\documentclass[a4paper, amsfonts, amssymb, amsmath, reprint,  nofootinbib, twoside,superscriptaddress]{revtex4-1}
\usepackage[english]{babel}
\usepackage[utf8]{inputenc}
\usepackage[colorinlistoftodos, color=green!40, prependcaption]{todonotes}
\usepackage[pdftex, pdftitle={Article}, pdfauthor={Author}]{hyperref} 
\begin{document}
\title{Surface exceptional points in a topological Kondo insulator}

\author{Robert Peters}
    \email[Correspondence email address: ]{peters@scphys.kyoto-u.ac.jp}
    \affiliation{Department of Physics, Kyoto University, Kyoto 606-8502, Japan}
\author{Kazuhiro Kimura}
    \affiliation{Department of Physics, Kyoto University, Kyoto 606-8502, Japan}      
    \author{Yoshihiro Michishita}
    \affiliation{Department of Physics, Kyoto University, Kyoto 606-8502, Japan}  
    \author{Tsuneya Yoshida}
    \affiliation{Department of Physics, University of Tsukuba, Ibaraki 305-8571, Japan}  
\author{Norio Kawakami}
    \affiliation{Department of Physics, Kyoto University, Kyoto 606-8502, Japan}    

\date{\today} 

\begin{abstract}
Correlated materials have appeared as an arena to study non-Hermitian effects as typically exemplified by the emergence of exceptional points.
We show here that topological Kondo insulators are an ideal platform for studying these phenomena due to strong correlations and surface states exhibiting a nontrivial spin texture. Using numerical simulations, we demonstrate the emergence of exceptional points in the single-particle Green's function on the surface of the material while the bulk is still insulating.
We reveal how quasiparticle states with long lifetimes are created on the surface by non-Hermitian effects while the Dirac cones are smeared, which explains the surface Kondo breakdown at which heavy Dirac cones disappear from the single-particle spectrum and are replaced by light states. 
We further show how the non-Hermiticty changes the spin texture inherent in the surface states, which might help identify exceptional points experimentally. Besides confirming the existence of non-Hermitian effects on the surface of a topological Kondo insulator, this paper demonstrates how the eigenstates and eigenvalues of the effective non-Hermitian matrix describing the single-particle Green's function help understand the properties of correlated materials.
\end{abstract}


\maketitle

\section{Introduction}

Recently, it has become apparent that non-Hermitian phenomena\cite{00018732.2021.1876991,PhysRevLett.120.146402,El-Ganainy2018} can be observed in single- or two-particle properties of correlated materials at equilibrium \cite{PhysRevB.98.035141,10.1093/ptep/ptaa059,Rausch_2021,PhysRevLett.121.026403,PhysRevB.99.201107,PhysRevB.101.085122,PhysRevLett.125.227204,Kozii_Fu,PhysRevLett.121.026403}.
Ordinarily, non-Hermiticity has been connected to systems out-of-equilibrium experiencing gain and loss, e.g., photonic systems, open quantum systems\cite{Feng2017,El-Ganainy2018,Mirieaar7709}, mechanical systems\cite{PhysRevLett.125.118001,PhysRevB.100.054109}, and active matter\cite{PhysRevLett.123.205502,Sone2020}, leading to exciting phenomena, such as the emergence of exceptional points\cite{Heiss_2012,PhysRevB.99.041406}. 
Exceptional points are topologically protected band touchings, where the eigenvalues (energies) and the corresponding eigenvectors of the non-Hermitian Hamiltonian are equal. 
Furthermore, experiments have demonstrated the emergence of several spectacular properties at or close to exceptional points\cite{Zhou_2018,Ozdemir2019}, such as loss-induced transparency, unidirectional invisibility\cite{PhysRevLett.106.213901,Regensburger2012,Feng2013}, topological chirality\cite{PhysRevLett.86.787,Gao2015,Doppler2016,Xu2016,Yoon2018}, band merging\cite{Zhen2015}, and enhanced sensitivity\cite{Hodaei2017,PhysRevLett.112.203901,PhysRevLett.117.110802,Chen2017}.

In correlated materials at equilibrium, non-Hermitian properties emerge due to the finite lifetime of quasiparticles in the single- or two-particle Green's functions\cite{PhysRevB.98.035141,10.1093/ptep/ptaa059,Rausch_2021,PhysRevLett.121.026403,PhysRevB.99.201107,PhysRevLett.125.227204,PhysRevB.103.125145,PhysRevB.99.121101,PhysRevB.100.115124,JPSJ.90.074703,PhysRevB.100.245205,PhysRevLett.126.176601}. Because Green's functions describe the material's response to an external perturbation and can be directly measured in experiments, the emergence of exceptional points  is expected to affect experimental observations. For example, the single-particle Green's function can be observed in angle-resolved photoemission spectroscopy (ARPES) and tunneling experiments. It has been shown that exceptional points induce peaks in the observed spectral function. Furthermore, exceptional points have also been found in two-particle Green's functions, e.g., resulting in the appearance of Fermi arcs in the charge-structure factor \cite{Rausch_2021}. Notably, it has been demonstrated that the non-Hermitian Hamiltonian describing an open quantum system with loss and gain and the effective non-Hermitian Hamiltonian describing the single-particle Green's function are equivalent\cite{PhysRevLett.124.196401}. Thus, it should be possible to observe similar spectacular phenomena in the single-particle Green's function of strongly correlated materials as in photonic and open quantum systems.

It has been shown that exceptional points can easily emerge in band structures of correlated materials hosting Dirac cones, where two noninteracting bands coalesce \cite{PhysRevB.98.035141,PhysRevB.99.201107}. In this situation,  a small difference in the lifetime of the quasiparticles of these two bands will lead to a splitting of the Dirac cone and the emergence of two exceptional points connected by a Fermi arc. Even small correlation effects can induce this behavior. Thus, correlated systems hosting a Dirac point seem to be particularly interesting when studying non-Hermitian properties and exceptional points. One such type of material hosting Dirac cones and being correlated is that of a topological Kondo insulator. Here, Dirac cones emerge on the surface of the material. Furthermore, the Dirac cones are composed of weakly interacting conduction ($c$) electrons and strongly correlated $f$ electrons. Thus, the lifetimes of the bands forming the Dirac cones are different, satisfying the desired condition for exceptional points to appear.
It has been shown that correlation effects are often stronger on the surface than in bulk due to the reduced coordination number \cite{PhysRevLett.114.177202,PhysRevB.93.235159}.
Thus, one might expect that exceptional points emerge at the surface while the bulk is still insulating, which would make the observation easy.  
Moreover, Dirac cones on the surfaces of a 3D Kondo insulator exhibit a spin texture.  This spin texture will be influenced by the exceptional points and can be used to detect them in experiments. 
Topological Kondo insulators thus seem to provide an exciting playground to study the emergence of exceptional points and the interplay between non-Hermitian properties induced by correlations and band topology\cite{PhysRevX.9.041015,RevModPhys.93.015005,PhysRevX.8.031079,PhysRevResearch.1.012003,Denner2021}.

In this paper, we study non-Hermitian properties in a three-dimensional topological Kondo insulator. The noninteracting model hosts Dirac cones on each surface at $\vec k=(0,0)$, $\vec k=(\pi,0)$, and $\vec k=(0,\pi)$ of the surface Brillouin zone (BZ). Using dynamical mean-field theory (DMFT), we calculate a self-consistent self-energy of this model in a wide range of temperatures and study the emergence of exceptional points in the bulk and on the surface. We confirm that correlation effects are enhanced on the surface compared to the bulk, leading to the emergence of exceptional points pinned to the surface while the bulk is still insulating. However, contrary to previous studies \cite{PhysRevB.98.035141,PhysRevB.99.201107}, the Dirac cones themselves do not split and do not form exceptional points. Instead, we find that exceptional points are formed by hybridization between the states comprising the Dirac cones and other surface states. 
Finally, we analyze the impact of non-Hermiticity and exceptional points 
on the surface band structure and the surface spin texture.

This paper is organized as follows: We introduce the model and the methods in the next section. This is followed by Sec. \ref{sec_bulk} describing our analysis of the bulk properties of this model. In Sec. \ref{sec_surface}, we show how non-Hermitian properties affect the surface band structure of the topological Kondo insulator, demonstrating the existence of exceptional points in the single-particle Green's function on the surface of the topological Kondo insulator. Finally, we summarize and conclude the paper.

\section{Model and Methods}
\label{sec_method}
\subsection{Model Hamiltonian}
To study the impact of non-Hermiticity on the single-particle properties of a topological Kondo insulator, we use the following model introduced in Refs.~\cite{PhysRevB.98.075104,PhysRevB.100.085124}
\begin{eqnarray}
H&=&H_0+H_{\mathrm{int}}\;,\\
H_0&=&\sum_{\vec k}\sum_{\tau=\{\uparrow,\downarrow\}}\sum_{o=\{c,f\}}\epsilon^o_{\vec k}c^\dagger_{\vec k,\tau,o}c_{\vec k,\tau,o}\nonumber\\
&&+V\sum_{\vec k,\tau_1,\tau_2}c^\dagger_{\vec k,\tau_1,c}c_{\vec k,\tau_2,f}\sin k_x\sigma^x_{\tau_1\tau_2}\nonumber\\
&&+V\sum_{\vec k,\tau_1,\tau_2}c^\dagger_{\vec k,\tau_1,c}c_{\vec k,\tau_2,f}\sin k_y\sigma^y_{\tau_1\tau_2}\nonumber\\
&&+V\sum_{\vec k,\tau_1,\tau_2}c^\dagger_{\vec k,\tau_1,c}c_{\vec k,\tau_2,f}\sin k_z\sigma^z_{\tau_1\tau_2}\nonumber\\
&&+E_c\sum_{i,\tau}n_{i,\tau,c}\;,\\
\epsilon^c_{\vec k}&=&-2t\Bigl(\cos(k_x)+\cos(k_y)+\cos(k_z)\Bigr)\nonumber\\
&&+4t'\cdot\cos(k_x)\cos(k_y)\nonumber\\&&+4t'\cdot\cos(k_y)\cos(k_z)\nonumber\\&&+ 4t'\cdot\cos(k_x)\cos(k_z)\nonumber\\
&&+8t''\cdot\cos(k_x)\cos(k_y)\cos(k_z)\;,\\
\epsilon^f_{\vec k}&=&-0.1\epsilon^c_{\vec k}\;,\nonumber\\
H_{\mathrm{int}}&=&U\sum_i n_{i,\uparrow,f}n_{i,\downarrow,f}\;.
\end{eqnarray}
The operator $c^\dagger_{\vec k,\sigma,o}$ creates an electron with momentum $\vec k$, spin direction $\sigma$ in orbital $o\in\{c,f\}$. $\epsilon^o_{\vec k}$ describes the energy depending on the momentum for each orbital.
$E_c$ is an energetical shift of the $c$ electrons.  We include nearest-neighbor, next-nearest neighbor, and next-next-nearest neighbor hoppings on a cubic lattice and take the nearest-neighbor hopping $t$ as a unit of energy throughout this paper. The model respects inversion and time-reversal symmetry.
We choose the parameters as $t'=t''=0.375t$ and $E_c=t$ resulting in a 
band structure similar to SmB$_6$. We find band inversions between the $c$ electrons and $f$ electrons at $(k_x, k_y,k_z)$=$(\pi,0,0)$, $(0,\pi,0)$, and $(0,0,\pi)$ in the Brillouin zone, identical to the situation in SmB$_6$. 
Due to the hybridization, $V$, between the $c$-electron band and the $f$-electron band, a gap opens in the bulk spectrum.Throughout this paper, we choose $V=0.8t$ and $U=16t$, which results in a strongly correlated insulating state at $T=0$. 
Thus, this model describes a strong topological Kondo insulator in 3D with inversion and time-reversal symmetry.  The sign of the hopping of the $c$ and the $f$ electrons should be opposite to open a topologically nontrivial gap.
We note that the results reported in this paper do not strongly depend on  the choice of parameters as long as the model remains in the strong topological insulating phase and the interaction strength is strong enough to form a crossover from an insulator to a metal due to the Kondo effect at finite temperature. Furthermore, a more realistic treatment of SmB$_6$ should also include the bands stemming from boron. The spectral function visible on the surface will then depend on the surface termination of the crystal\cite{PhysRevB.90.075131,JPSCP.3.017038,PhysRevB.104.075131}. While the described phenomena will occur, as long as surface states are present, exceptional points might be more difficult to detect experimentally because of the additional spectral weight of the boron bands.

At each momentum, the noninteracting model reads
\begin{eqnarray}
H_0&=&\sum_{\tau_1,\tau_2}\left(\epsilon^f c_{\tau_1,f}^\dagger c_{\tau_2,f}\delta_{\tau_1\tau_2}+\epsilon^c c_{\tau_1,c}^\dagger c_{\tau_2,c}\delta_{\tau_1\tau_2}\right.\\&&\left.+c_{\tau_1,f}^\dagger c_{\tau_2,c}\vec V\cdot \vec\sigma_{\tau_1\tau_2}+\text{h.c.}\right)\nonumber,
\end{eqnarray}
where $\vec V$ is a vector describing the hybridization between $c$ and $f$ electrons, and $\tau_1$, $\tau_2$ are spin indices.
By rotating the spin quantization axis of the $c$- and the $f$-electron band in the direction of $\vec V$, only $\sigma_z$ appears in the rotated Hamiltonian, and the Hamiltonian is block-diagonal. 
Thus, regardless of the exact momentum dependence of $\epsilon^f_{\vec k}$, $\epsilon^c_{\vec k}$, and $\vec V_{\vec k}$, this Hamiltonian can be block diagonalized.

\subsection{Dynamical Mean-Field Theory}
To include correlation effects and calculate a self-consistent self-energy, we use DMFT\cite{RevModPhys.68.13}. DMFT maps each lattice site onto a quantum impurity model by calculating the local Green's function. DMFT calculates a frequency-dependent self-energy. Furthermore, DMFT can be easily extended to real-space DMFT, where each atom of a finite cluster or a slab is mapped onto its impurity model. This enables one to study models with open boundaries. 
In Sec. \ref{sec_bulk}, we perform calculations using DMFT for the bulk, where the self-energy of each atom is the same. The local Green's function is then calculated via integration over the whole BZ. In Sec. \ref{sec_surface}, we use real-space DMFT to analyze a slab of $20$ layers with open boundary conditions \cite{PhysRevB.93.235159,PhysRevB.98.075104}. By integrating over the slab's two-dimensional (2D) BZ, we calculate a local Green's function for each layer, map each layer onto its impurity model, and calculate the layer-dependent self-energy.  Our results show that the self-energy is nearly layer-independent for layers $4$ to $16$. Therefore, we believe that the number of $20$ layers is enough to study correlation effects and the emergence of exceptional points on the surface of a topological Kondo insulator.

To calculate the self-energy of a quantum impurity model, we use the numerical renormalization group (NRG)\cite{RevModPhys.80.395}, which is well suited for calculating real-frequency spectral functions and self-energies at low temperatures with high resolution around the Fermi energy for arbitrary interaction strengths.
We note that we have used the same model and a combination of methods to study the magnetic properties of topological Kondo insulators\cite{PhysRevB.98.075104}. 

\subsection{Non-Hermitian properties in the single-particle Green's function}
Single-particle properties of a correlated system are given by the retarded single-particle Green's functions,
\begin{equation}
    G(\vec k,\omega)=\Bigl(\omega-H_0(\vec k)-\Sigma(\omega)\Bigr)^{-1},
\end{equation}
where $\vec k$ is the momentum, $\omega$ the frequency, $\eta\rightarrow 0$ a convergence factor, $H_0$ the noninteracting tight-binding Hamiltonian, and $\Sigma(\omega)$ the retarded self-energy. $H_0$ and $\Sigma(\omega)$ are thereby in general matrices. 
Using DMFT, the self-energy is a  momentum-independent diagonal matrix, ignoring nonlocal fluctuations. 
Thus, the Green's function can be written as
\begin{equation}
    G(\vec k,\omega)=\left(\omega+i\eta-H_{\mathrm{eff}}(\vec k,\omega)\right)^{-1},
\end{equation}
where $H_{\mathrm{eff}}(\vec k,\omega)=H_0(\vec k) + \Sigma(\omega)$ is an effective non-Hermitian Hamiltonian. 
The effective Hamiltonian is a non-Hermitian matrix because of the imaginary part of the self-energy corresponding to the lifetime of the particles.  It may happen that this matrix is defective at certain points $(\vec k,\omega)$, i.e., the effective Hamiltonian cannot be diagonalized. At these points, $(\vec k,\omega)$, at least two eigenvalues and eigenstates of the effective Hamiltonian coalesce. We call these points exceptional points. 

The eigenvalues and eigenvectors of this effective Hamiltonian completely describe the single-particle Green's function.
Away from the exceptional points, we can diagonalize the effective Hamiltonian as
\begin{eqnarray}
H_{\mathrm{eff}}\vert n_R\rangle&=&E_n\vert n_R\rangle,\\
\langle n_L\vert H_{\mathrm{eff}}&=&E_n\langle n_L\vert,\\
\langle n_L\vert m_R\rangle&=&\langle n_R\vert m_L\rangle=\delta_{nm},
\end{eqnarray}
where $\vert n_L\rangle$ and $\vert n_R\rangle$ are the left- and right-eigenvectors of the non-Hermitian matrix $H_{\mathrm{eff}}$, respectively.
$E_n$ is a complex-valued eigenvalue of the matrix.
Furthermore, $\langle n_L\vert=\left(\vert n_L\rangle\right)^H$ and $\langle n_R\vert=\left(\vert n_R\rangle\right)^H$ hold, where the superscript $H$ is the Hermitian adjoint.
Using these vectors, we can write the retarded Green's function as
\begin{eqnarray}
G_{ij}(\omega,\vec k)&=&\sum_n\langle i\vert n_R\rangle\frac{1}{\omega-E_n}\langle n_L\vert j\rangle,
\end{eqnarray}
where $i$ and $j$ are orbital and spin indices of the original model.
Note that the eigenvectors and the eigenvalues depend on the momentum and the frequency.

Besides being directly observable in ARPES and tunneling experiments, the single-particle Green's function also determines single-particle expectation values, such as the spin expectation value.
In this paper, we will analyze the impact of non-Hermiticity in the spectral function and the impact on the spin expectation values that can be observed in spin-resolved ARPES.

We can write the single-particle spectral function and  expectation values using the left- and right-eigenvectors of the non-Hermitian matrix $H_{\mathrm{eff}}$,
\begin{eqnarray}
A_{ij}(\omega,\vec k)&=& -\frac{1}{\pi}\text{Im}\left(\sum_n\langle i\vert n_R\rangle\frac{1}{\omega-E_n}\langle n_L\vert j\rangle\right),\nonumber\\ \\
\langle\sigma^{x,y,z}\rangle(\omega,\vec k)&=&-\frac{1}{\pi}\text{Im}\sum_{ij} \left(  \sigma^{x,y,z}_{ij}G_{ji}(\omega)\right),\nonumber\\
&=&-\frac{1}{\pi}\text{Im}\left(\sum_{ij} \sum_n\langle j\vert\sigma^{x,y,z}\vert i\rangle\right.\nonumber\\&&\quad\quad\quad\left.\times\langle i\vert n_R\rangle\frac{1}{\omega-E_n}\langle n_L\vert j\rangle\right),\nonumber\\
&=&-\frac{1}{\pi}\text{Im}\left( \sum_n\langle n_L\vert\sigma^{x,y,z}\vert n_R\rangle\right.\nonumber\\&&\quad\quad\quad\quad\quad\quad\times\left.\frac{1}{\omega-E_n}\right),
\end{eqnarray}
where $A_{ij}(\omega,\vec k)$ is the spectral function for orbitals $i$ and $j$ at frequency $\omega$ and momentum $\vec k$, and $\langle\sigma^{x,y,z}\rangle(\omega,\vec k)$ is the spin expectation value as observed in spin-resolved ARPES. $\sigma^{x,y,z}$ are the matrix representations of the spin operators for the noninteracting Hamiltonian $H_0$. We note that the eigenstates and eigenvalues depend via the effective Hamiltonian on the frequency $\omega$.
Because the eigenstates of the effective Hamiltonian determine the spin direction of a photoelectron in ARPES at a specific frequency, we will later show the spin expectation values of the eigenstates.

\section{bulk properties}
\label{sec_bulk}
\begin{figure}[t]
\includegraphics[width=\linewidth]{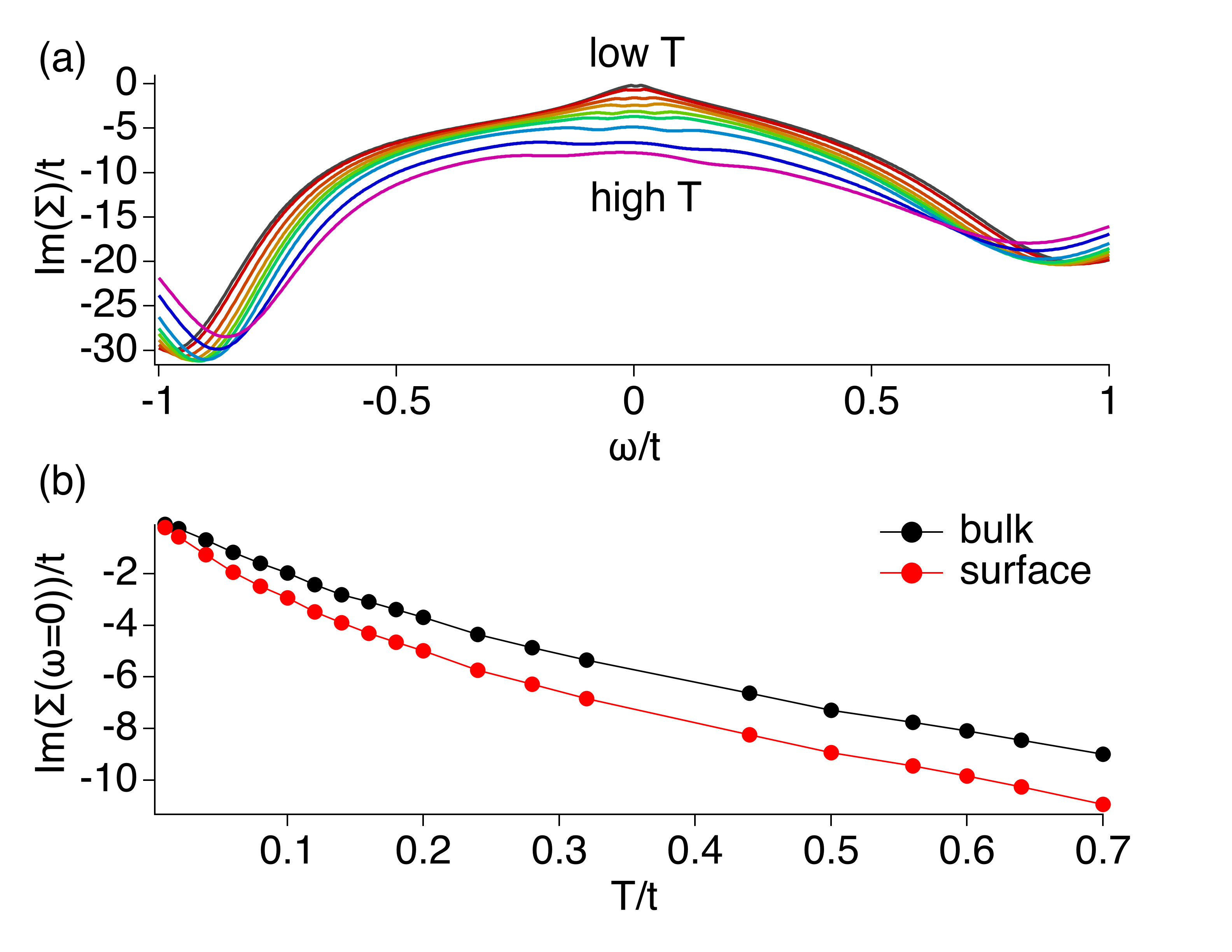}
\caption{  Imaginary part  of the bulk self-energy for different temperatures (a). Comparison between the imaginary part of the bulk and surface self-energy at $\omega=0$ (b). The self-energies in (a) are calculated at the following temperatures $T/t=\{0.02$, $0.04$, $0.08$, $0.12$, $0.16$, $0.2$, $0.28$, $0.44$, $0.56\}$.  \label{Fig1}}
\end{figure}
\begin{figure}[t]
\includegraphics[width=\linewidth]{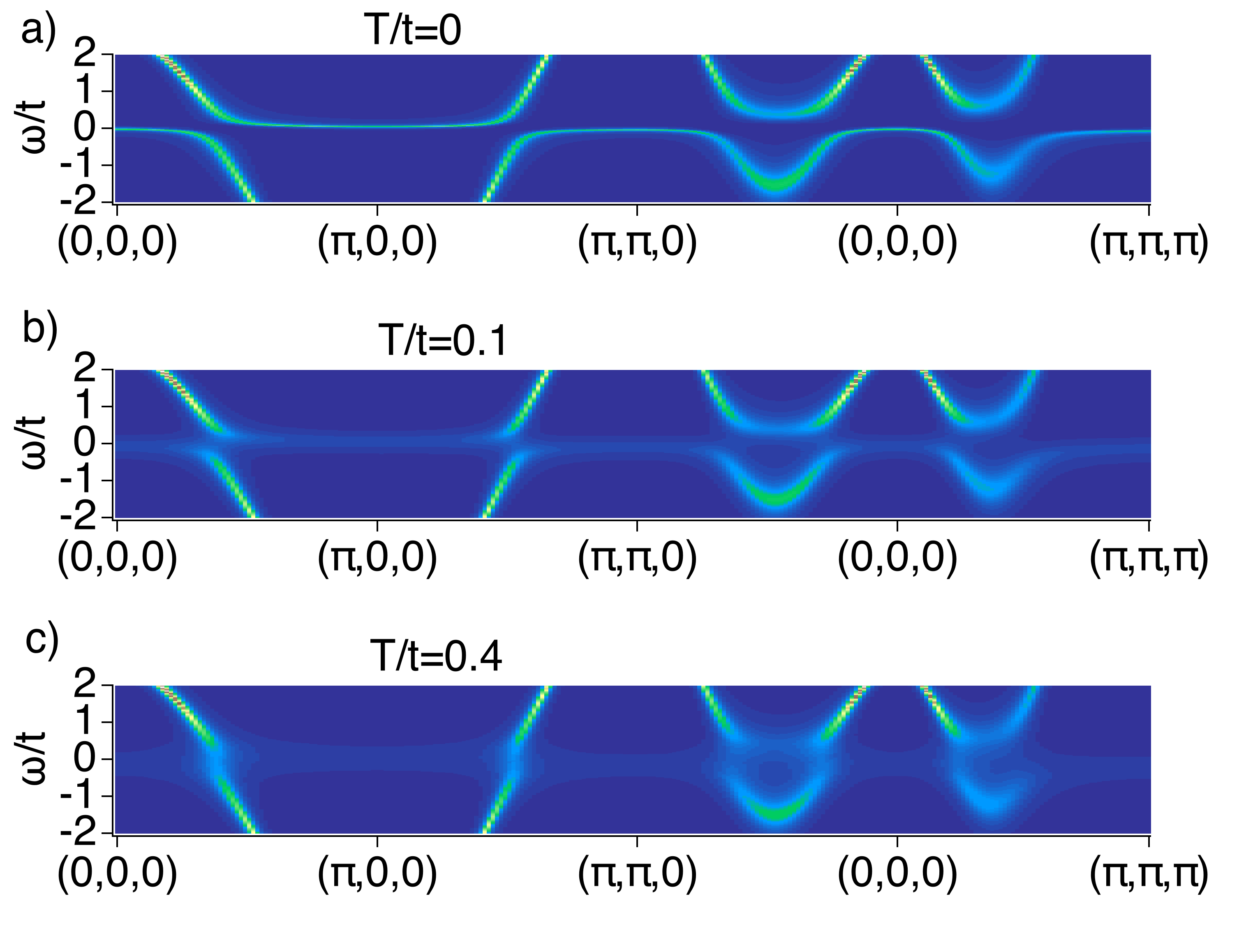}
\caption{Momentum-resolved spectral functions in the bulk for three different temperatures, $T/t=0$ (a), $T/t=0.1$ (b), and $T/t=0.4$ (c).
\label{Fig2}}
\end{figure}
\begin{figure*}[t]
\includegraphics[width=0.32\linewidth]{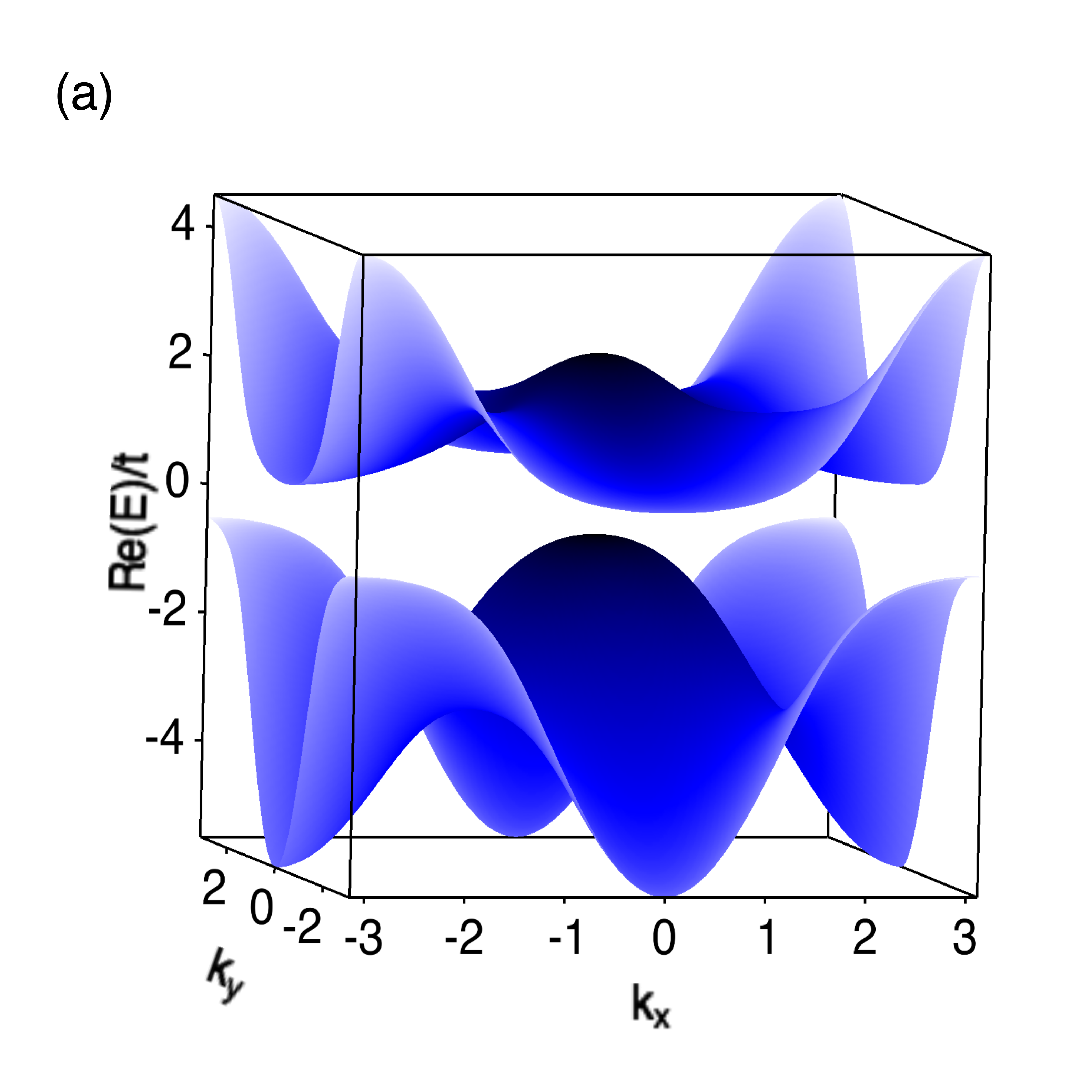}
\includegraphics[width=0.32\linewidth]{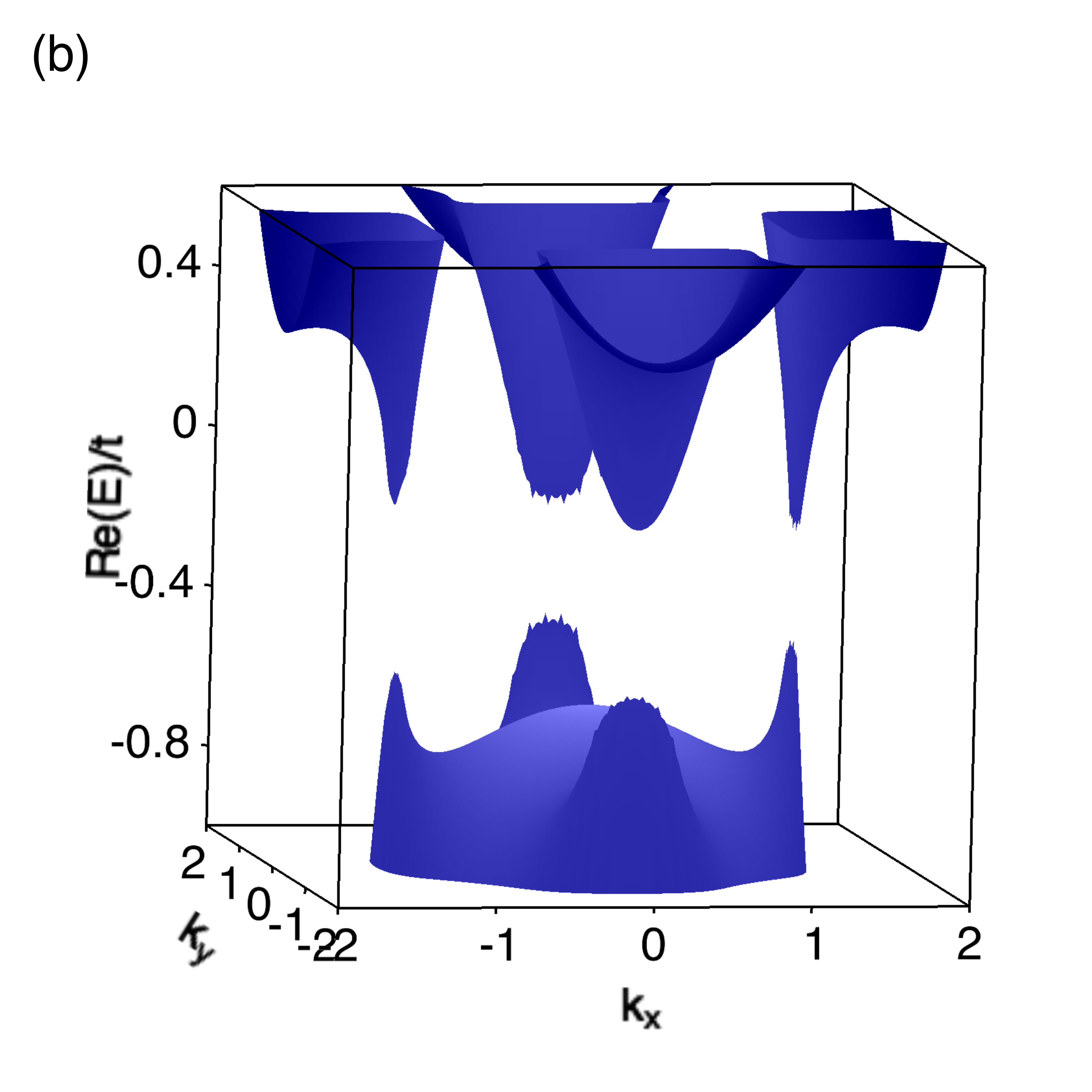}
\includegraphics[width=0.32\linewidth]{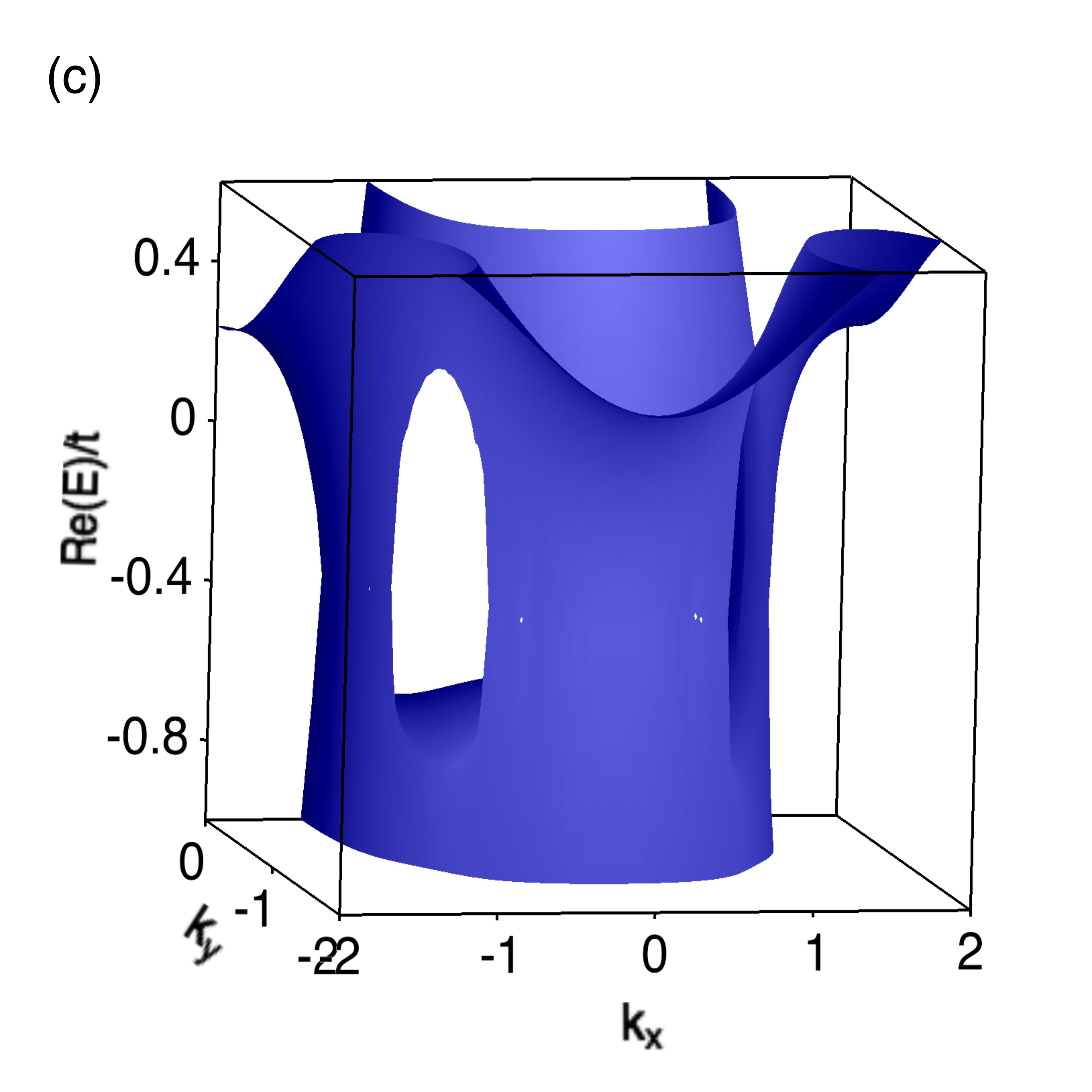}
\caption{Real part of the eigenvalues over the momentum of the effective Hamiltonian, $H_{\mathrm{eff}}(k_x,k_y,k_z=0,\omega=0)$, in the bulk for $T/t=0.0$ (a), $T/t=0.16$ (b), and $T/t=0.2$ (c).  We show only a part of the BZ in (b) and (c) for better visibility of the non-Hermitian effect.  \label{Fig3}}
\end{figure*}
\begin{figure}[t]
\includegraphics[width=\linewidth]{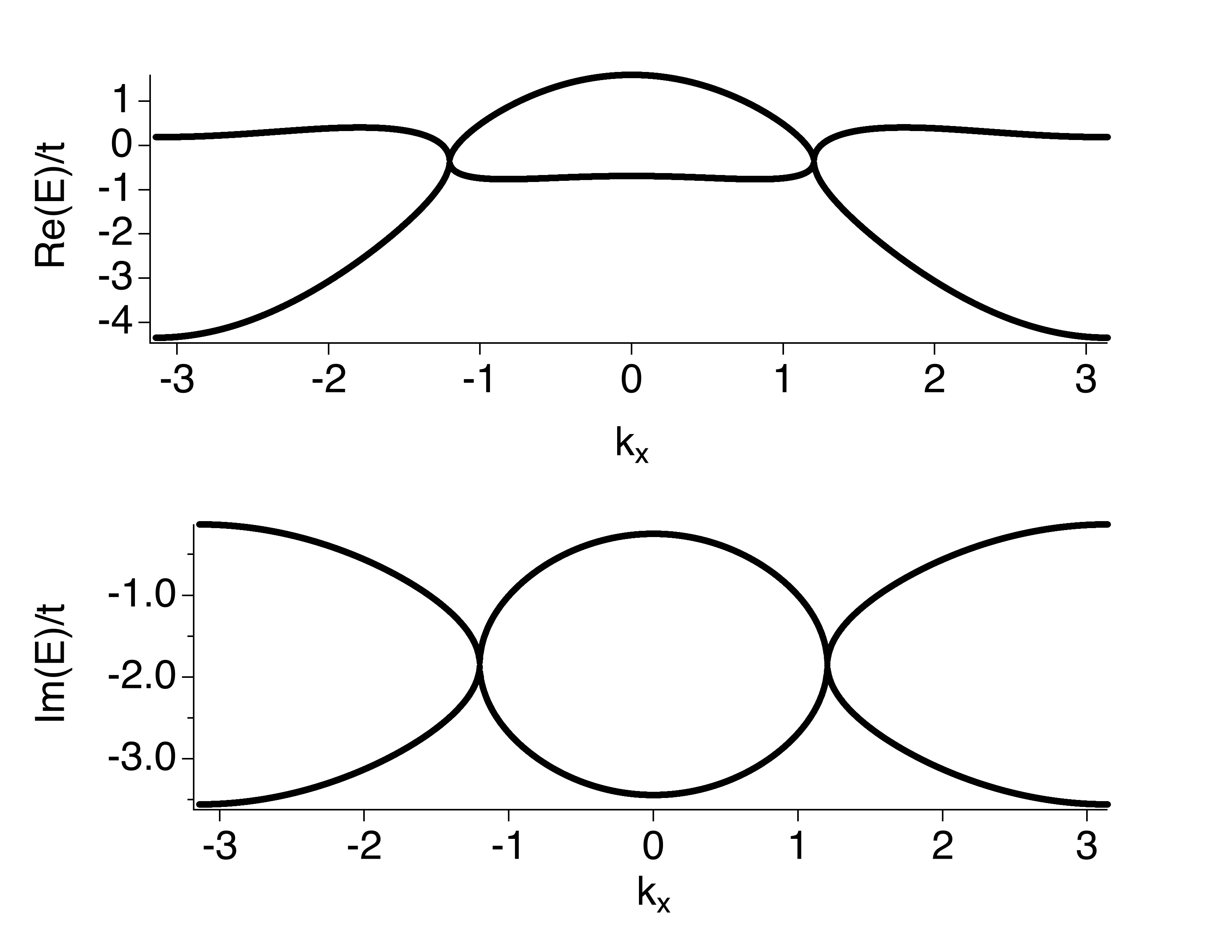}
\caption{Real and imaginary part of the eigenvalues of the effective Hamiltonian for $\omega=0$ and $k_y=0.74$ and $k_z=0$ for different $k_x$ at $T/t=0.2$. This line of momenta goes through two exceptional points in the BZ.
\label{Fig4}}
\end{figure}
In this section, we analyze the bulk properties of this model  at finite temperatures.  In particular, we demonstrate the appearance of exceptional points in the bulk spectrum around the Kondo temperature at which the system changes from insulating to metallic, which has also been shown for other $f$-electron systems in Ref.~\cite{PhysRevB.101.085122,PhysRevLett.125.227204}.
 We will show in Sec.~\ref{sec_surface} that exceptional points emerge at the surface of the material at a lower temperature than in the bulk.

The imaginary part of the bulk self-energy for different frequencies and a comparison between the bulk and the surface self-energy at  $\omega=0$ for different temperatures are shown in Fig. \ref{Fig1}. 
This imaginary part, which corresponds to an inverse lifetime of the quasiparticles, leads to the emergence of non-Hermitian properties in the single-particle Green's function. 
While the imaginary part vanishes at $\omega=0$ for $T=0$, it becomes large at finite temperatures. Furthermore, we see that the imaginary part of the self-energy is larger in magnitude on the surface than in the bulk, indicating that the surface is more strongly correlated than the bulk. This has also been found in a previous study \cite{PhysRevB.93.235159}.

The effective Hamiltonian at the Fermi energy completely describes the single-particle properies of the material at the Fermi energy.
We thus mainly focus on the properties of the effective Hamiltonian at  $\omega=0$.
Because the imaginary part of the self-energy vanishes at $T=0$ at the Fermi energy, the effective Hamiltonian at $T=0$ is just given by the noninteracting Hamiltonian, where energy levels are possibly shifted by the real part of the self-energy. In the current model, the hybridization between the $c$ and $f$ electrons leads to a gap at $\omega=0$ in the noninteracting Hamiltonian in the bulk. Thus, the effective Hamiltonian is gapped, and the system is insulating, as shown in Fig.~\ref{Fig2}.
At high temperatures, on the other hand, the imaginary part of the self-energy becomes large. 
For a two-band model consisting of a noninteracting $c$-electron band and a correlated $f$-electron band, we can calculate the eigenvalues of the effective Hamiltonian in the limit of a very large imaginary part as
\begin{eqnarray}
H_{\mathrm{eff}}(\vec k,\omega=0)&=&\begin{pmatrix}\epsilon^c_k&V_k\\V_k^\star&\epsilon^f_k+\Sigma(\omega=0)\end{pmatrix}\\
\Rightarrow (E_1,E_2)&=&\left(\epsilon^c_k,\epsilon^f_k+\Sigma(\omega=0)\right)\quad\text{for Im}\Sigma\rightarrow\pm\infty.\nonumber
\end{eqnarray}
If the imaginary part of the self-energy becomes very large, the effect of the hybridization becomes negligible, and we find a unperturbed $c$ electron and a correlated $f$ electron. Thus, at high temperatures, when the imaginary part of the self-energy is large, the system becomes metallic with a Fermi surface given by the conduction electrons.

The single-particle spectrum of the bulk consisting of four bands is shown in Fig.~\ref{Fig2} for three different temperatures. 
At $T=0$, the single-particle spectrum is gapped in the bulk. Because the imaginary part of the self-energy vanishes, $f$ and $c$ electrons can hybridize at the Fermi energy, $\omega=0$, and thus form a gap.
With increasing temperature, the crossover from the insulating state at $T=0$ to a metallic state at high temperatures occurs. At high temperatures, $f$ electrons localize and thus cannot hybridize with the $c$ electrons at the Fermi energy. Thus, the $c$ electron bands span the gap, and the system becomes metallic. 

Recently, it has become clear that exceptional points appear in the single-particle spectrum close to the Fermi energy approximately around the Kondo temperature \cite{PhysRevB.101.085122}. Thus, we here analyze the energy eigenvalues of the effective Hamiltonian at the Fermi energy, $\omega=0$. Figure \ref{Fig3} shows the real part of the eigenvalues of the effective Hamiltonian for $k_z=0$ over $(k_x,k_y$). We see the existence of a gap at $\omega=0$. Because these eigenvalues completely determine the spectral function at $\omega=0$, we confirm that the system is insulating. Note that the gap size in Fig.~\ref{Fig3} appears larger than that in Fig.~\ref{Fig2}. The gap in Fig.~\ref{Fig2} appears smaller because the effective Hamiltonian changes with the frequency.

Figure~\ref{Fig3}(b) shows the eigenvalues of the effective Hamiltonian for $T/t=0.16$ close to the Kondo temperature. While $H_0$ is the same as in Fig.~\ref{Fig3}(a), the self-energy has changed. The imaginary part of the $f$-electron self-energy  results in the formation of "pockets" where the eigenvalues from the band below and above the Fermi energy approach each other. The momenta of these pockets are given by the minima of the hybridization along the lines of $\epsilon^c_{\vec k}=\epsilon^f_{\vec k}=0$ in the noninteracting Hamiltonian.
Further increasing the temperature and thus the imaginary part in the self-energy leads to a merging of the pockets from below and above the Fermi energy, as shown in Fig. \ref{Fig3}(c). This merging leads to a line of momenta where the real part of the eigenvalues of the effective Hamiltonian is degenerate. Furthermore, at the edge of the pockets, the real and imaginary parts of two eigenvalues are equal, corresponding to an exceptional point in the effective Hamiltonian. The line inside the merged pockets, where the real parts of two eigenvalues are the same, is called a Fermi arc. 

Increasing the temperature above the Kondo temperature, the size of the pockets increases, and finally, pockets developing at different momenta merge. At this temperature, also the exceptional points merge and annihilate each other \cite{PhysRevB.98.035141}.

To demonstrate the existence of these exceptional points, we show the eigenvalues of the bulk Hamiltonian along a straight line through the exceptional points at $k_z=0$ in Fig. \ref{Fig4}. We see that exactly at the same momentum, the real part and the imaginary part of the eigenvalues are the same. Thus, all bands are degenerate in the effective Hamiltonian at this point. We note that this cannot be an ordinary (Hermitian) degeneracy because the hybridization between the bands does not vanish at this momentum. Thus, this degeneracy originates in the non-Hermiticty induced by the imaginary part of the self-energy.

\section{Non-Hermitian effects on the surface states}
\label{sec_surface}
\begin{figure}[t]
\includegraphics[width=\linewidth]{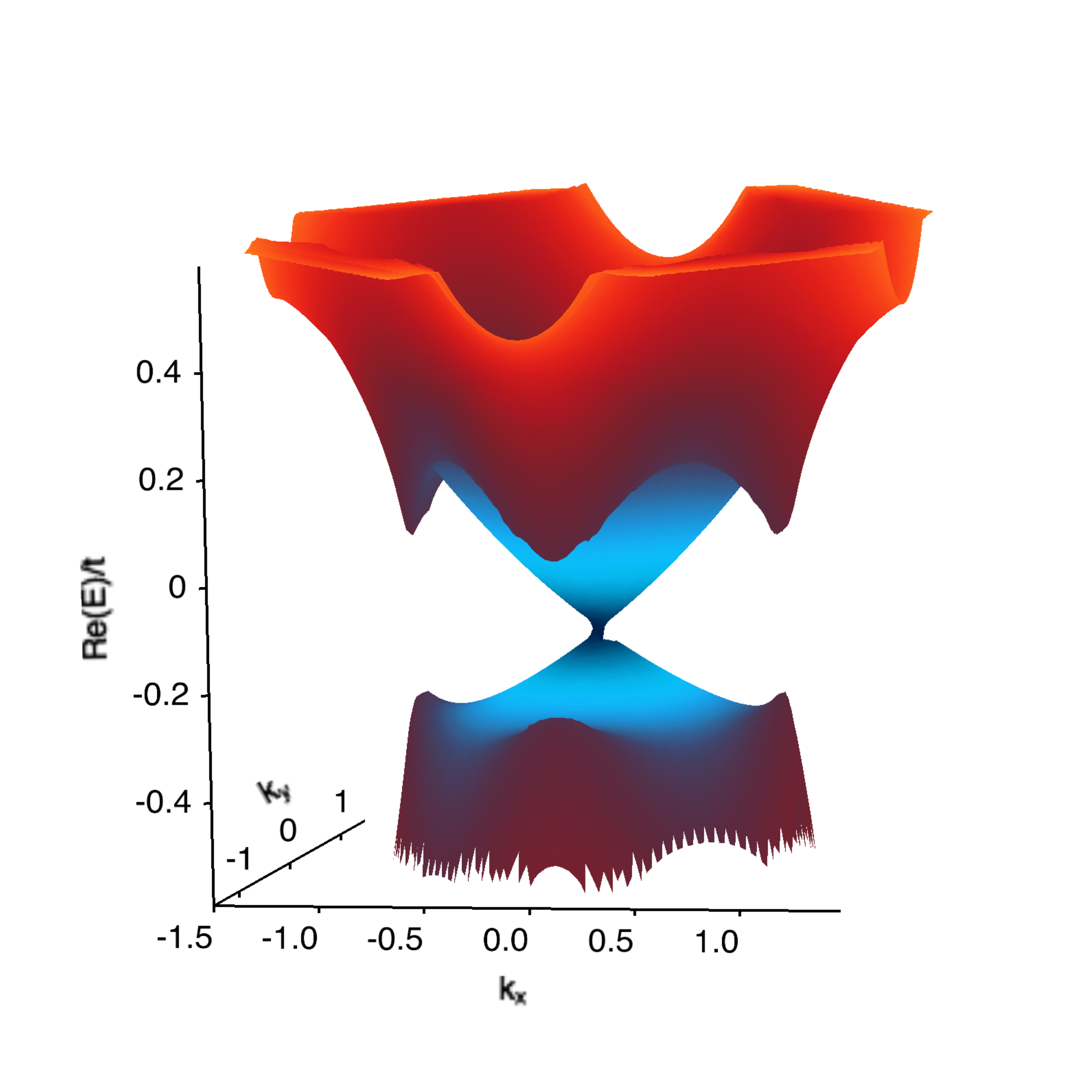}
\caption{Real part of the eigenvalues of the effective Hamiltonian ($T/t=0.106$) for an open system consisting of $20$ layers focusing on $k_x\in[-1.5,1.5]$ and $k_y\in[-1.5,1.5]$ of the surface BZ. The bulk is still gapped at this temperature. Pockets in the band structure are visible at $(k_x,k_y)\approx(0,-0.9)$, $(-0.9,0)$, and $(0.9,0)$. The color is added for better visibility and corresponds to the distance from the origin.
\label{Fig5}}
\end{figure}
\begin{figure}[t]
\includegraphics[width=\linewidth]{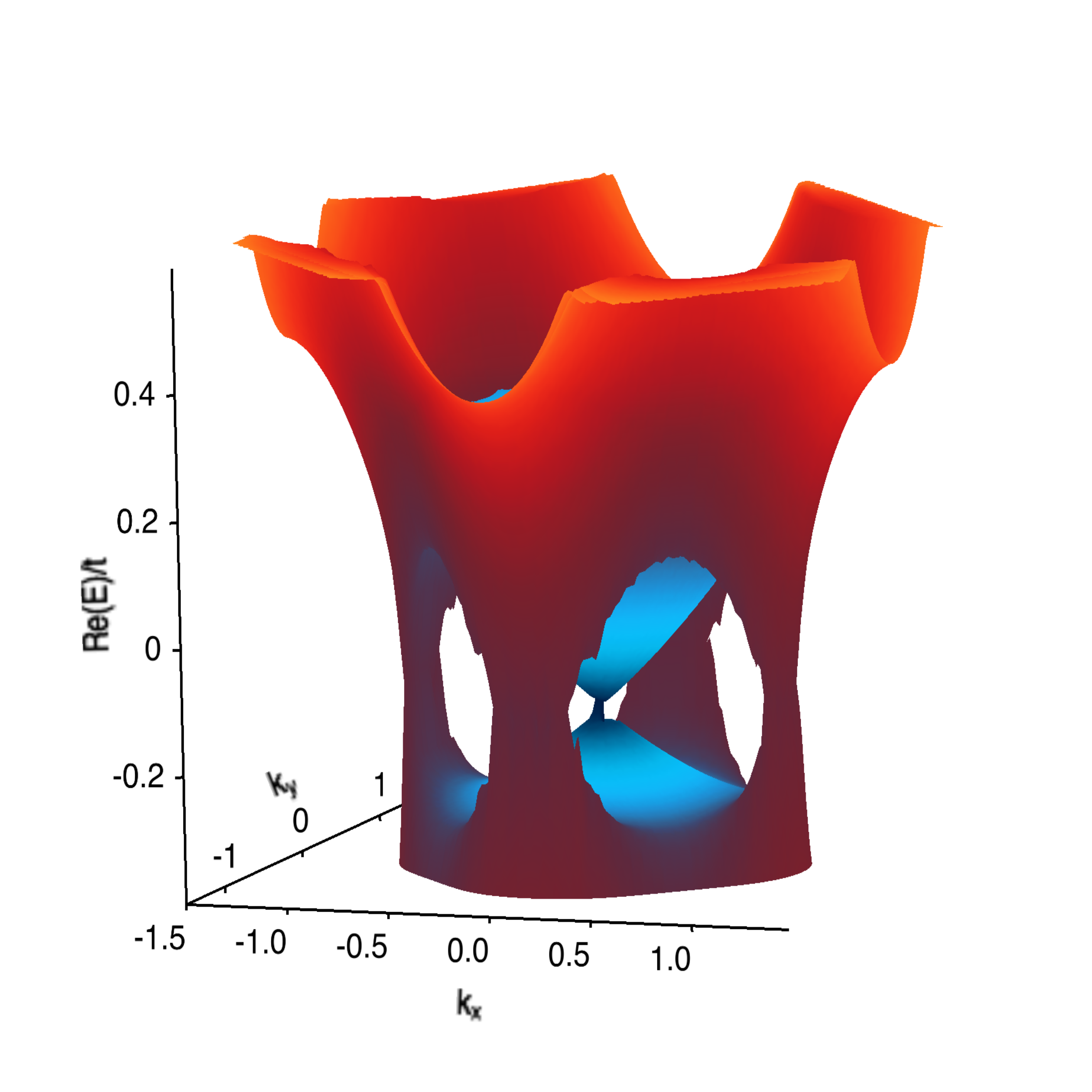}
\caption{ Same as Fig.~\ref{Fig5} but for $T/t=0.11$.\label{Fig6}}
\end{figure}
\subsection{Energy spectrum of $H_{\mathrm{eff}}$}
We now turn to the main results of this paper, i. e., how non-Hermiticity affects the surface states of a topological Kondo insulator.
Previous calculations have shown that even a small difference in the lifetime of the particles forming a Dirac cone can result in the splitting of the Dirac cone and the formation of exceptional points connected by Fermi arcs \cite{PhysRevB.98.035141}.
Thus, we next focus on the effect of non-Hermiticity on the surface states of this topological Kondo insulator. Due to the topology of this system, there are three protected Dirac cones on each surface with surface momentum $(k_x,k_y)=(0,0)$, $(\pi,0)$, and $(0,\pi)$. 
Furthermore, because the magnitude of the imaginary part of the self-energy at the surface is enhanced compared to the bulk, as shown in Fig. \ref{Fig1}, non-Hermitian effects can be expected to be stronger on the surface than in the bulk.
We here ask whether the above scenario can apply for the surface Dirac states in the present system and how else non-Hermiticity affects the surface spectrum.

In Fig.~\ref{Fig5}, we show the real part of the eigenvalues for a system with open boundaries in the $z$-direction consisting of $20$ layers at $T/t=0.106$. We here focus on the Dirac cone at the center of the surface BZ. We find a similar behavior for the Dirac cones located at $(k_x,k_y)$=$(\pi,0)$ and $(0,\pi)$.

The structure of the eigenvalues in Fig.~\ref{Fig5} shows two notable features: a Dirac cone at $(k_x,k_y)$=$(0,0)$ and pockets in the eigenvalue structure of the effective Hamiltonian separated from the Dirac cone.
The first observation is that the Dirac cone on the surface of this topological Kondo insulator is not strongly affected by the finite lifetime of the $f$ electrons at this temperature. This is contrary to the results in Ref.~\cite{PhysRevB.98.035141}, where the Dirac cone in the bulk of a 2D system splits by forming exceptional points.
This difference can be explained in the following way: There is no effective hybridization between the lower and the upper part of the Dirac cone on the same surface. Analyzing the current model, we find that there would be a hybridization between the Dirac cone on the bottom surface and that on the top surface, but there is no hybridization between the two bands comprising the Dirac cone on the same surface. The hybridization between states on two different surfaces is too small to affect the results because of the distance between the surfaces. Thus, exceptional points cannot emerge within the Dirac cone because of the absence of hybridization between the surface states. 
The second notable feature is the appearance of pockets in the eigenvalue structure. These pockets, which are absent at $T=0$, are created by the non-Hermiticity similar to Fig.~\ref{Fig3}(b) in the bulk at $T/t=0.16$.
The pockets in the eigenvalue spectrum with open boundaries appear at a much lower temperature than in the bulk, which can be explained by larger correlation effects on the surface.

With increasing the temperature to $T/t=0.11$, as shown in Fig.~\ref{Fig6}, we see that the pockets from above and below the Fermi energy merge. The emergent structure points to the existence of exceptional points at the edges of the pockets, which can also be confirmed by showing the concrete level structure (see Fig.~\ref{Fig8}) and analyzing the eigenvectors of the effective Hamiltonian. Thus, exceptional points emerge on the surface of the topological Kondo insulator for much smaller temperatures and do not split the Dirac cone. 
\begin{figure}[t]
\includegraphics[width=\linewidth]{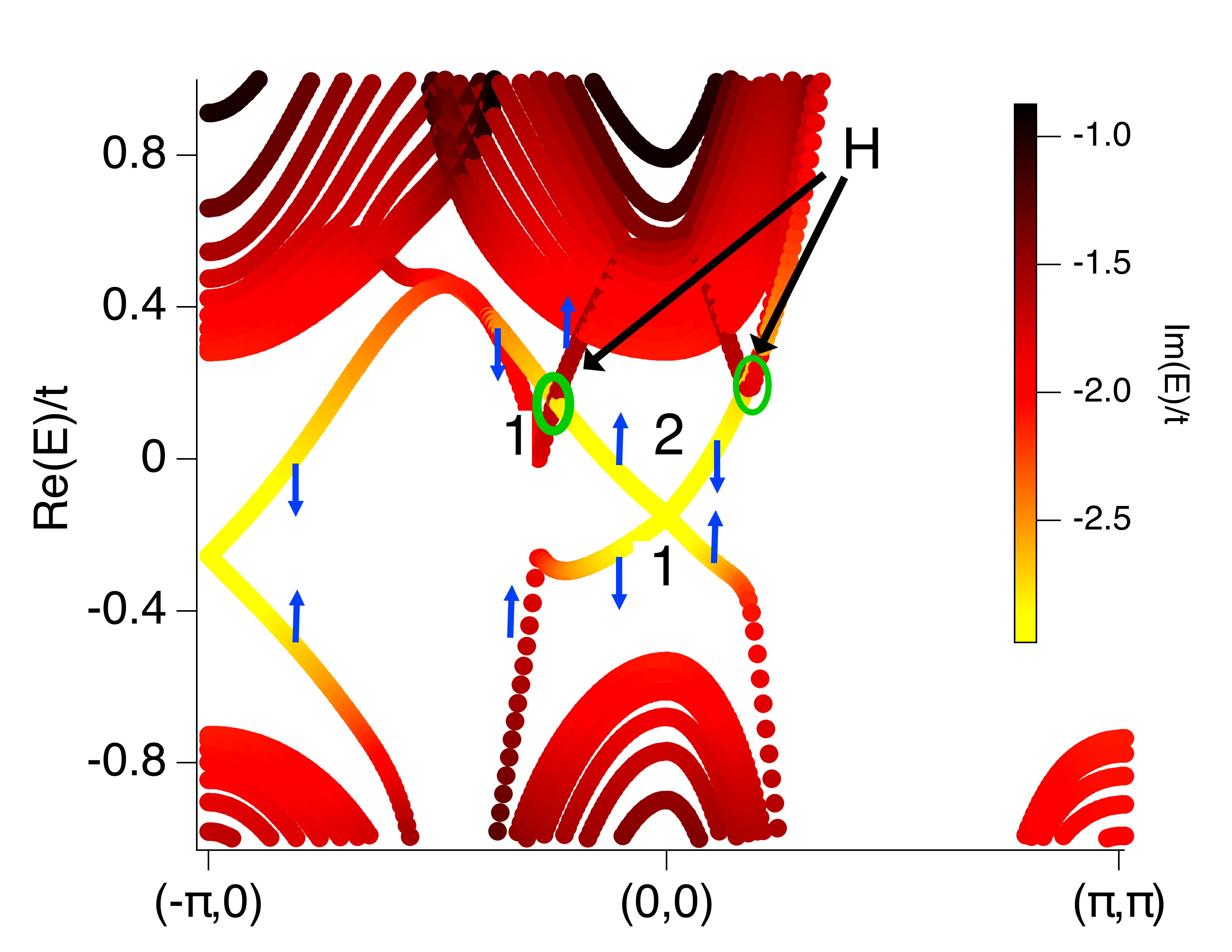}
\caption{Real part of the eigenvalues of the effective Hamiltonian at $\omega=0$  and $T/t=0.106$.
The color encodes the imaginary part of the eigenvalues. Furthermore, we include indices ($1$ or $2$) corresponding to the block of the block-diagonalized Hamiltonian, and mark degeneracies due to the absence of any hybridization by green circles and an 'H'. We also include the spin in the $y$-direction, $\langle \sigma^x\rangle$, of the surface states located on the bottom of the slab by blue arrows.\label{Fig7}}
\end{figure}
\begin{figure}[t]
\includegraphics[width=\linewidth]{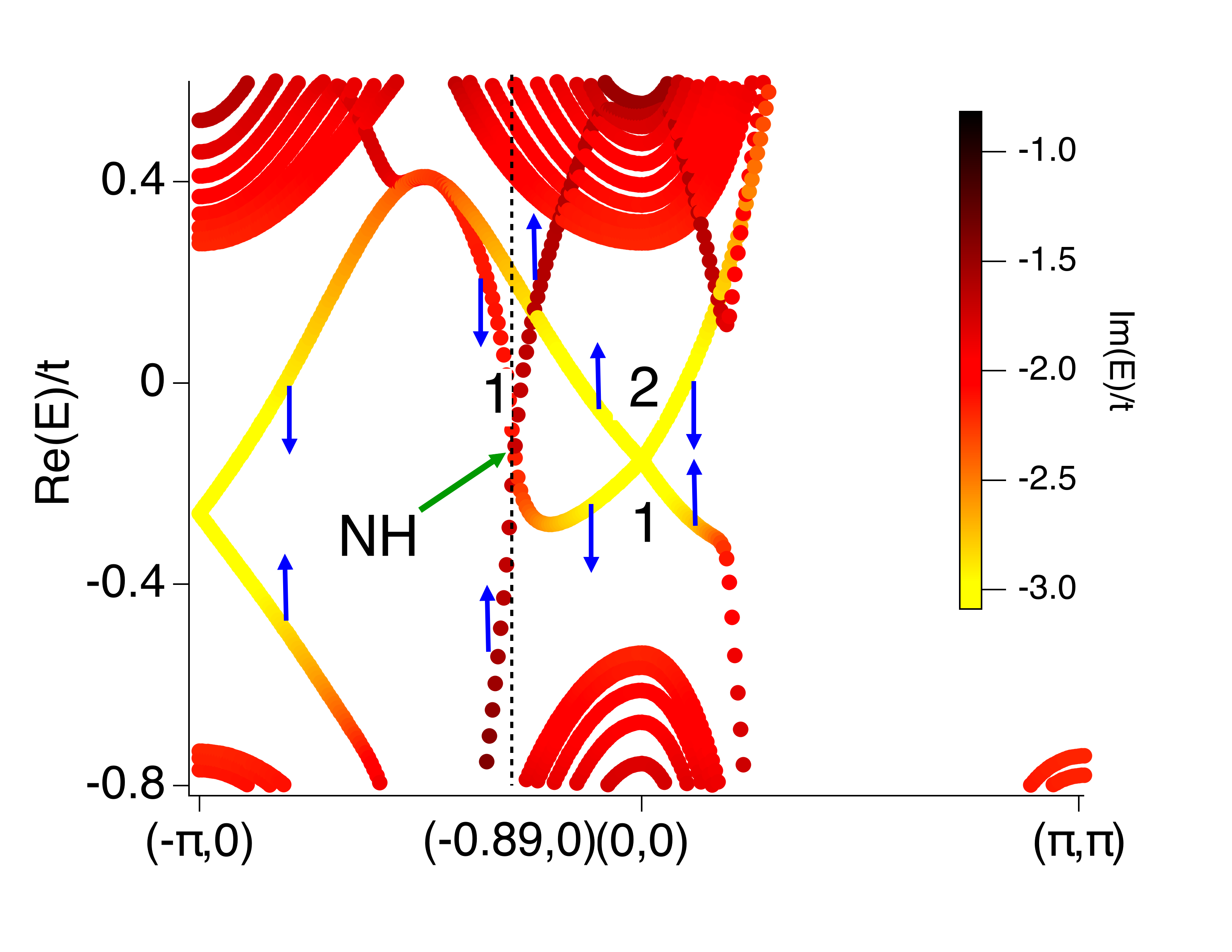}
\caption{Same as Fig.~\ref{Fig7} but for $T/t=0.11$.
We mark a degeneracy in the real part due to non-Hermiticity by 'NH'. 
\label{Fig8}}
\end{figure}
\begin{figure}[t]
\includegraphics[width=\linewidth]{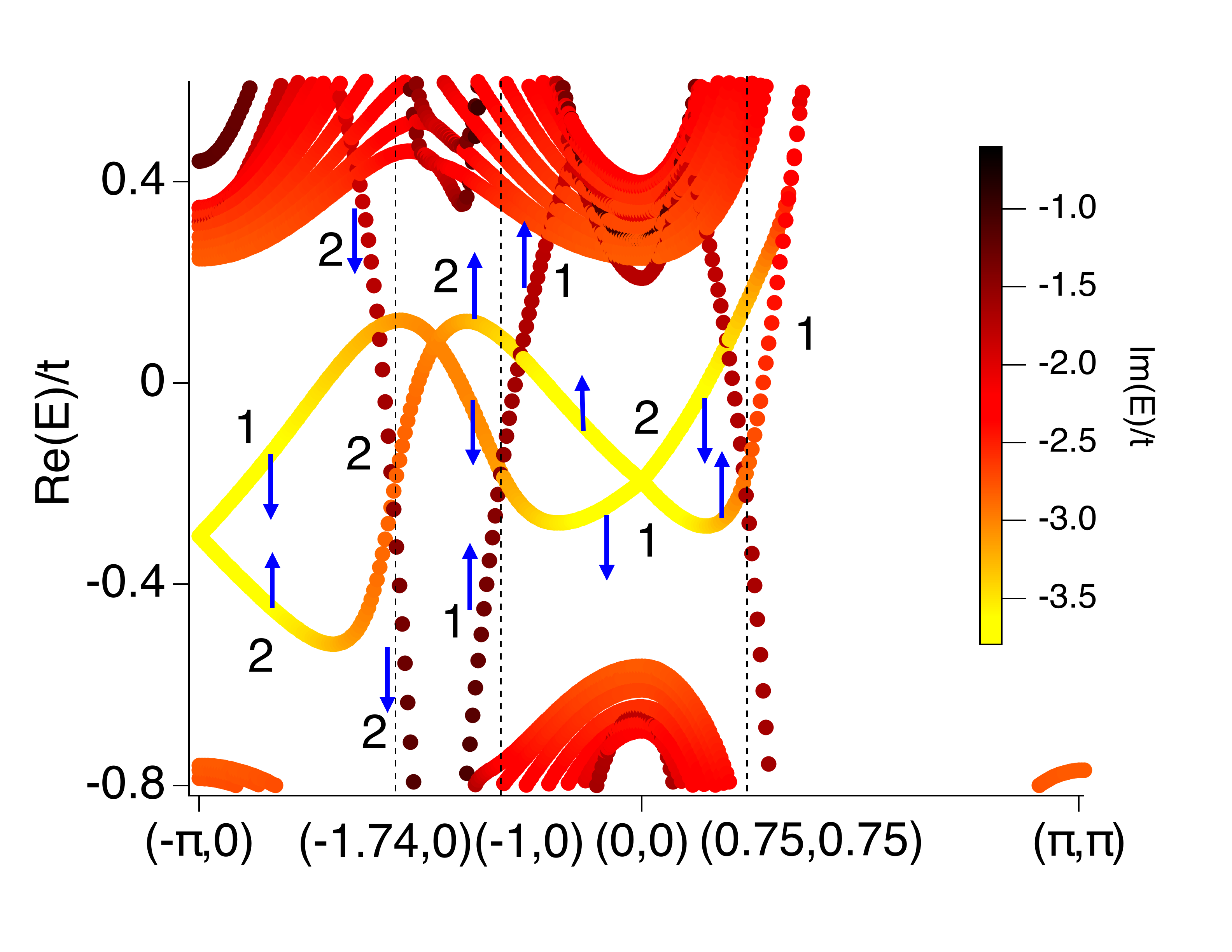}
\caption{Same as Fig.~\ref{Fig7} but for $T/t=0.14$.
\label{Fig9}}
\end{figure}

\subsection{Analysis of the degeneracies in the spectrum}
A more detailed analysis of the eigenvalues including the spin expectation values of the eigenstates is shown in Figs. \ref{Fig7}-\ref{Fig9}.
We show the real part of the eigenvalues close to the Fermi energy of the effective Hamiltonian at $\omega=0$ for different temperatures along the surface momentum from $(k_x,k_y)$=$(-\pi,0)$ via $(0,0)$ to $(\pi,\pi)$. Besides showing the real part of the eigenvalues, we also include the imaginary part as color.
As explained above, the Hamiltonian can always be block-diagonalized. To better understand the effect
of non-Hermiticity, we thus block-diagonalize the Hamiltonian and include an index corresponding to the block in Figs.~\ref{Fig7}-\ref{Fig9}.
While bands with the same index lie in the same block, bands with different indices lie in separate blocks and thus do not couple. 
When taking into account energies of both blocks, all bands are degenerate. However, because of inversion and time-reversal symmetry in this model, degenerate surface states lie on opposite surfaces.
Thus, by block-diagonalization, we can uniquely identify states on a specific surface of the topological insulator.
Finally, we also include the spin expectation value in the $y$-direction, $\langle\sigma^y\rangle$, of these states as blue arrows.

In Fig.~\ref{Fig7} at $T/t=0.106$, which is the same temperature as in Fig.~\ref{Fig5}, we see several band crossings.
We see the Dirac cones at $(k_x,k_y)$=$(0,0)$ and $(-\pi,0)$ due to the topological properties.
Analyzing the bands forming the Dirac cone at $(k_x,k_y)$=$(0,0)$ on the bottom surface of the slab, we see that both bands lie in different blocks of the block-diagonalized Hamiltonian. Thus, there is no hybridization between these states in the noninteracting and the effective Hamiltonian. As a hybridization between states is essential to form exceptional points, this explains why the Dirac cones do not split and do not form exceptional points in the presence of correlations, although they are made of particles with different lifetimes. 
We note that the possibility to block-diagonalize the Hamiltonian is completely general for a four-band model (consisting of two spinful bands) with inversion and time-reversal symmetry, as explained in the model section.

Around $(k_x,k_y)$=$(-0.9,0)$ in Fig.~\ref{Fig7}, we see the appearance of the pockets at this temperature. The pockets are created by the band forming the lower part (block index $1$) of the Dirac cone at $(k_x,k_y)$=$(0,0)$, and a band originating from the bulk spectrum at $(0,0)$. 
We see that these bands have the same block index and thus hybridize with each other.
Furthermore, we see that the band, forming the pocket above the Fermi energy at $(k_x,k_y)$=$(-0.9,0)$, crosses the band comprising the upper part of the Dirac cone at $(k_x,k_y)$=$(0,0)$ without hybridization. By analyzing the block indices of these bands, we see that these bands lie in different blocks and thus do not hybridize. In Fig.~\ref{Fig7}, we have marked Hermitian degeneracies due to the absence of hybridization by green circles and an 'H'.

In Fig.~\ref{Fig8}, at $T/t=0.11$, the pockets from above and below the Fermi energy have merged and formed a point where the real parts of two eigenvalues are degenerate, denoted as 'NH' in this figure.  Note that this non-Hermitian degeneracy is absent on the line from $(k_x,k_y)$=$(0,0)$ to $(\pi,\pi)$. Thus, when going around the center of the BZ, this line of real-part degeneracies has to end at some momenta, as visible in Fig.~\ref{Fig6}. These endpoints must be exceptional points. We have verified these exceptional points by confirming that the real and imaginary parts of the eigenstates are degenerate at these endpoints.
We thus find exceptional points on the surface of the material emerging due to correlation effects.
Further analyzing the imaginary part of the eigenstates forming the degeneracy at $(k_x,k_y)$=$(-0.9,0)$, we see that the imaginary parts of these bands are smaller in magnitude than that of the states comprising the Dirac cone at $(k_x,k_y)$=$(0,0)$, which will become important further below.

At $T/t=0.11$ in Fig.~\ref{Fig8}, only the band contributing to the lower part of the Dirac cone at $(k_x,k_y)$=$(0,0)$ forms a non-Hermitian degeneracy. With increasing the temperature further to $T/t=0.14$, as shown in Fig.~\ref{Fig9}, the band contributing to the upper part of the Dirac cone also forms a non-Hermitian degeneracy at $(k_x,k_y)$=$(-1.74,0)$.
While there is a degeneracy of the real part on the line of momenta from $(k_x,k_y)$=$(-\pi,0)$ to $(0,0)$ in the band with block index 2, it is absent on the line of momenta from $(k_x,k_y)$=$(0,0)$ to $(\pi,\pi)$, which points again to the existence of exceptional points. Besides these non-Hermitian degeneracies, there are several new Hermitian degeneracies between bands that do not hybridize.

\subsection{Effect of the non-Hermiticity on the momentum-dependent spectral weight}

\begin{figure}[t]
\includegraphics[width=\linewidth]{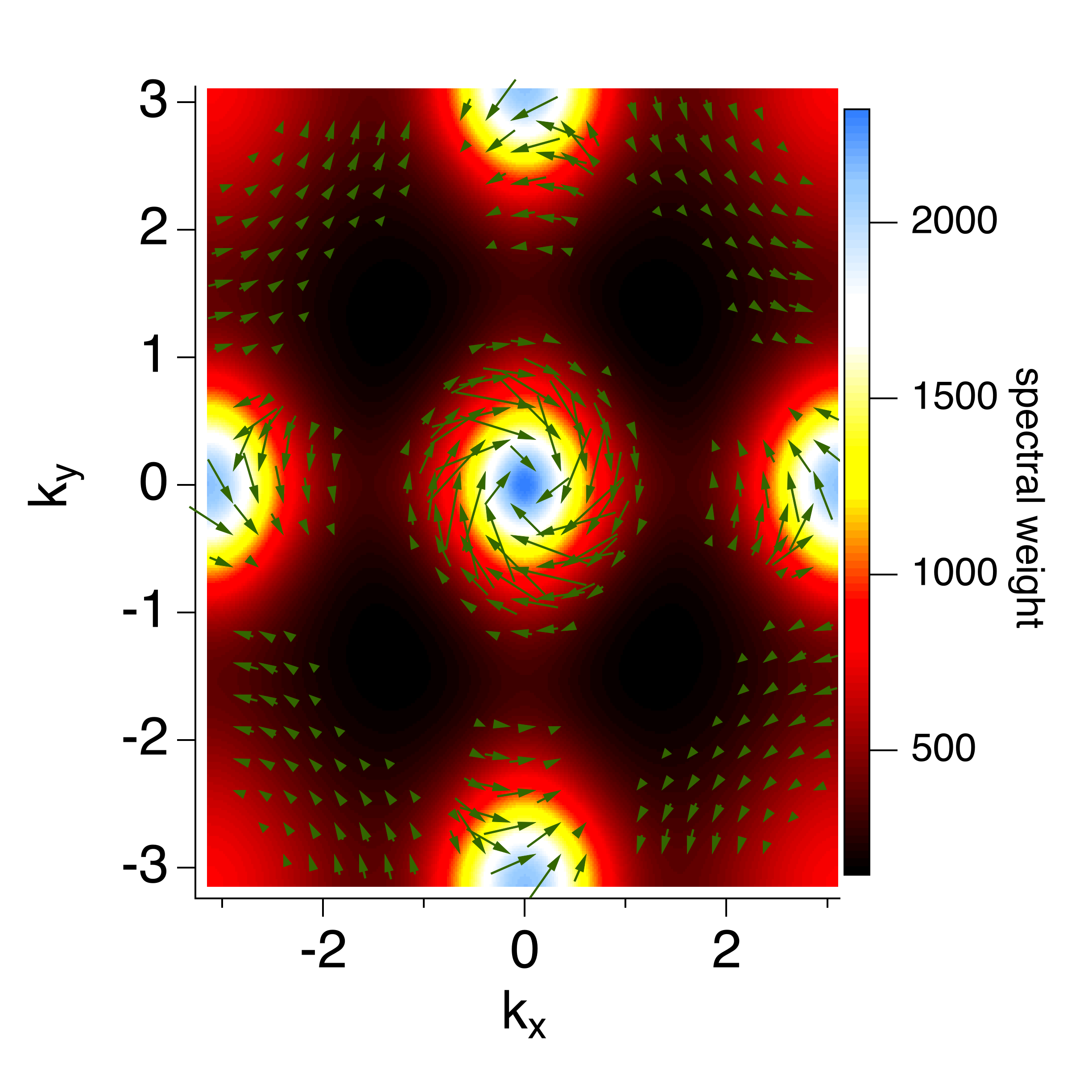}
\caption{Surface spectral function for $\omega=0$ and the direction of the spin expectation values on the bottom of the slab as calculated from the Green's function for $T/t=0.01$.  \label{Fig10}}
\end{figure}
\begin{figure}[t]
\includegraphics[width=\linewidth]{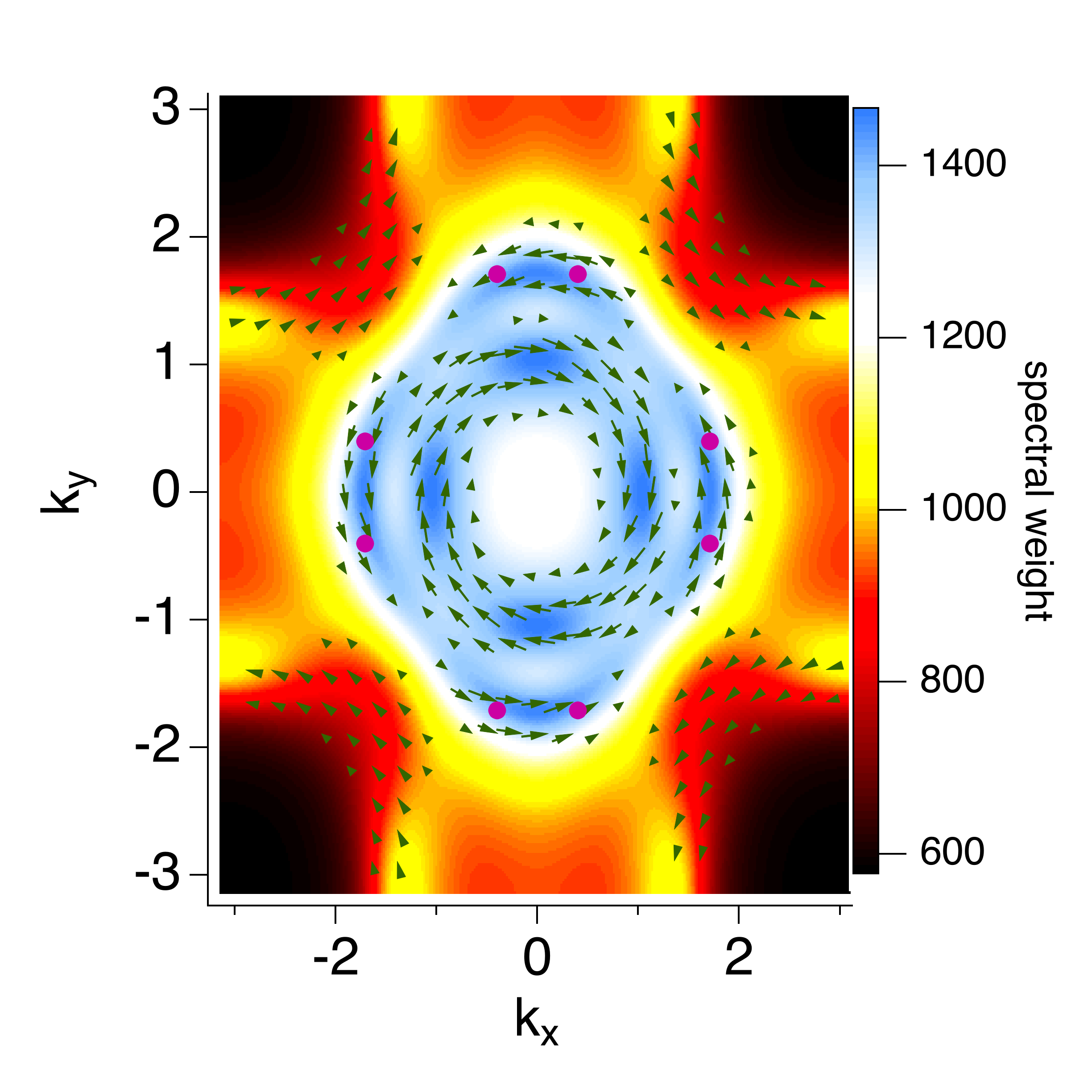}
\caption{Same as Fig.~\ref{Fig10} but for $T/t=0.14$. We show the location of the exceptional points as red dots. We stress the two ring-like structures in the spectrum, visible as dark blue color, going around the origin of the BZ.
\label{Fig11}}
\end{figure}

Finally, let us analyze the non-Hermitian effect directly on the single-particle Green's function as observable in ARPES. We show the spectral function of the single-particle Green's function at the Fermi energy for different temperatures in Fig.~\ref{Fig10} and \ref{Fig11}. We also include the spin expectation value as calculated from the Green's function. However, we only take into account electrons on the bottom surface of the slab by summing only over the lower half of the slab 
\begin{equation}
    \langle\vec \sigma\rangle=-\frac{1}{\pi}\text{Im}\sum_{ij<N/2}\vec \sigma_{ij}G_{ji}(\omega=0),
\end{equation}
where $i$ and $j$ are indices describing the orbital, spin, and layer of the system. For a system consisting of $N$ layers, we only sum over layers in the lower half of the system. We note that the spin expectation value of the full system vanishes as the top surface and the bottom surface have opposite spin expectation values.

In Fig.~\ref{Fig10}, we show the spectral function at $T/t=0.01$. Clearly visible is the spectral weight created by the Dirac cones at $(k_x,k_y)$=$(0,0)$, $(\pi,0)$, and $(0,\pi)$. Focusing on the Dirac cone at $(k_x,k_y)$=$(0,0)$, we see that the spin rotates clockwise around the origin. Comparing with Fig.~\ref{Fig7}, we see that the Dirac cone is located slightly below the Fermi energy. The Fermi energy is located in the band forming the upper part of the Dirac cone, as shown in Fig.~\ref{Fig7}. The spin expectation value of this band (shown as an arrow in Fig.~\ref{Fig7}) exhibits a positive $\langle\sigma_y\rangle$ value for $k_x<0$ and a negative value for $k_x>0$, which explains the clockwise spin direction around $(k_x,k_y)$=$(0,0)$. This band mainly contributes to the spin expectation value seen in Fig.~\ref{Fig10}.

With increasing the temperature to $T/t=0.14$, as shown in Fig.~\ref{Fig11}, the momenta with highest spectral weight (blue color) shift away from $(k_x,k_y)$=$(0,0)$. This is an interesting feature as we see in Fig.~\ref{Fig9} that the position of the Dirac cone does not change with increasing temperature. 
However, Fig.~\ref{Fig9} reveals the existence of bands crossing the Fermi energy at $(k_x,k_y)$=$(-1.74,0)$, $(-1,0)$, and $(0.75,0.75)$ due to non-Hermitian effects. Furthermore, the imaginary part of the bands forming the Dirac cone is much larger in magnitude than the imaginary part of the bands crossing the Fermi energy away from $(k_x,k_y)$=$(0,0)$.
 Thus, the spectral weight of the Dirac cone is smaller than that of the  bands crossing the Fermi energy away from $(k_x,k_y)$=$(0,0)$.
These bands create two rings around the center of the BZ in the spectral function, as shown in Fig.~\ref{Fig11}.
In Fig.~\ref{Fig9}, we see that these bands have a large velocity compared to the bands comprising the Dirac cone.
 The situation described here is identical to the phenomenon called surface Kondo breakdown in Refs.~\cite{PhysRevLett.114.177202,PhysRevB.93.235159}. The heavy Dirac cone (small velocity) becomes invisible in the single-particle spectral function and is replaced by bands with large velocity. By analyzing the effective non-Hermitian Hamiltonian, we understand that the Dirac cone  still exists at $(k_x,k_y)$=$(0,0)$ but is smeared out due to a large imaginary part. The light states are created by a non-Hermitian effect suppressing the hybridization between different surface states.

Using the eigenstates of the effective non-Hermitian Hamiltonian describing the Green's function, as shown in Fig.~\ref{Fig9}, we can also understand the spin expectation values in Fig.~\ref{Fig11}.
The spin direction of the band with the smaller imaginary part (longest lifetime) causing the degeneracy at $(k_x,k_y)$=$(-1,0)$ rotates clockwise, 
 and the spin direction of the band with the smaller imaginary part creating the degeneracy at $(k_x,k_y)$=$(-1.74,0)$ rotates anticlockwise, as shown in Fig.~\ref{Fig9}.
 We thus understand the emergence of the two rings of high spectral weight in Fig.~\ref{Fig11}, where the spin direction rotates clockwise for the inner ring and anticlockwise for the outer ring.

Finally, as noted above, the degeneracy in the real part that exists in the band with index $2$ between $(k_x,k_y)=(-\pi,0)$ and $(0,0)$ is absent on the line between $(k_x,k_y)=(0,0)$ and $(\pi,\pi)$. Thus, there are line segments in the BZ that end in exceptional points.
In Fig.~\ref{Fig11}, we show these exceptional points as dark red dots.  These exceptional points lie on the surface of the material. Furthermore, because the surface is more strongly correlated than the bulk, the bulk is still insulating and does not form non-Hermitian degeneracies.  As expected from the analysis in Fig.~\ref{Fig9}, these points lie in the outer circle going around the center of the BZ. This ring exhibits variations of spectral weight depending on the momentum. These variations originate in the changes of the imaginary part of the band and the sudden change of the band structure at the exceptional points at which the band degeneracy suddenly vanishes.
Furthermore, close to these exceptional points, the spin direction visible in the outer ring becomes very small and seems to end. 
We believe that these variations of spectral weight and the spin expectation values yield the necessary clues to identify exceptional points in the single-particle Green's function of correlated materials.

\section{Conclusions}
\label{sec_conclusions}
We have analyzed non-Hermitian properties of a topological Kondo insulator in 3D. Using numerical calculations, we have confirmed that the crossover from localized $f$ electrons at high temperatures to hybridized $f$ electrons at low temperatures is a non-Hermitian effect accompanied by the appearance of exceptional points in the single-particle Green's function \cite{PhysRevB.101.085122}. Because correlation effects are stronger on the surface than in the bulk, exceptional points emerge on the surface while the bulk is still insulating. We found that the surface Dirac cones are stable against non-Hermitian effects, although comprised of particles with different lifetimes. This stability can be explained by the absence of hybridization between the states comprising the Dirac cone in this time-reversal and inversion symmetric system. Thus, exceptional points on the surface of the material form away from the Dirac cone by hybridization between different surface states. Notably, the emergence of exceptional points located on the surface is related to the surface Kondo breakdown investigated in previous studies \cite{PhysRevLett.114.177202,PhysRevB.93.235159}. While the states forming the Dirac cone acquire a large imaginary part and are smeared out, two new bands with a small imaginary part are created due to the non-Hermiticity of the effective Hamiltonian. These bands become visible in the single-particle spectral function. 
 Furthermore, we have analyzed the spin texture arising due to the topological surface states and non-Hermiticity and found that the surface states inherited from the non-Hermiticity have opposite spin directions.  

In summary, we believe that topological Kondo insulators host an ideal platform to study non-Hermitian effects in correlated materials. These materials provide exceptional points at the surface while the bulk is insulating, and the states due to non-Hermiticity exhibit a spin texture that can be used to detect those states.
Finally, our analysis based on the effective non-Hermitian Hamiltonian has been shown to be a powerful tool to detect and understand changes in the single-particle spectral function, such as the surface Kondo breakdown.

\begin{acknowledgments}
RP would like to thank Youichi
Yanase for fruitful discussions.
This work is supported by the WISE program,
MEXT, and by the following JSPS KAKENHI grants No. 20J12265,
No. JP18K03511, No. JP19H01838, No. JP21K13850, and No.~JP20H04627. Computer simulations were done on
the supercomputer of Tokyo University at the ISSP.
\end{acknowledgments}

\appendix
\section{Frequency dependence of the exceptional points}
\begin{figure*}[tbh]
\includegraphics[width=0.32\linewidth]{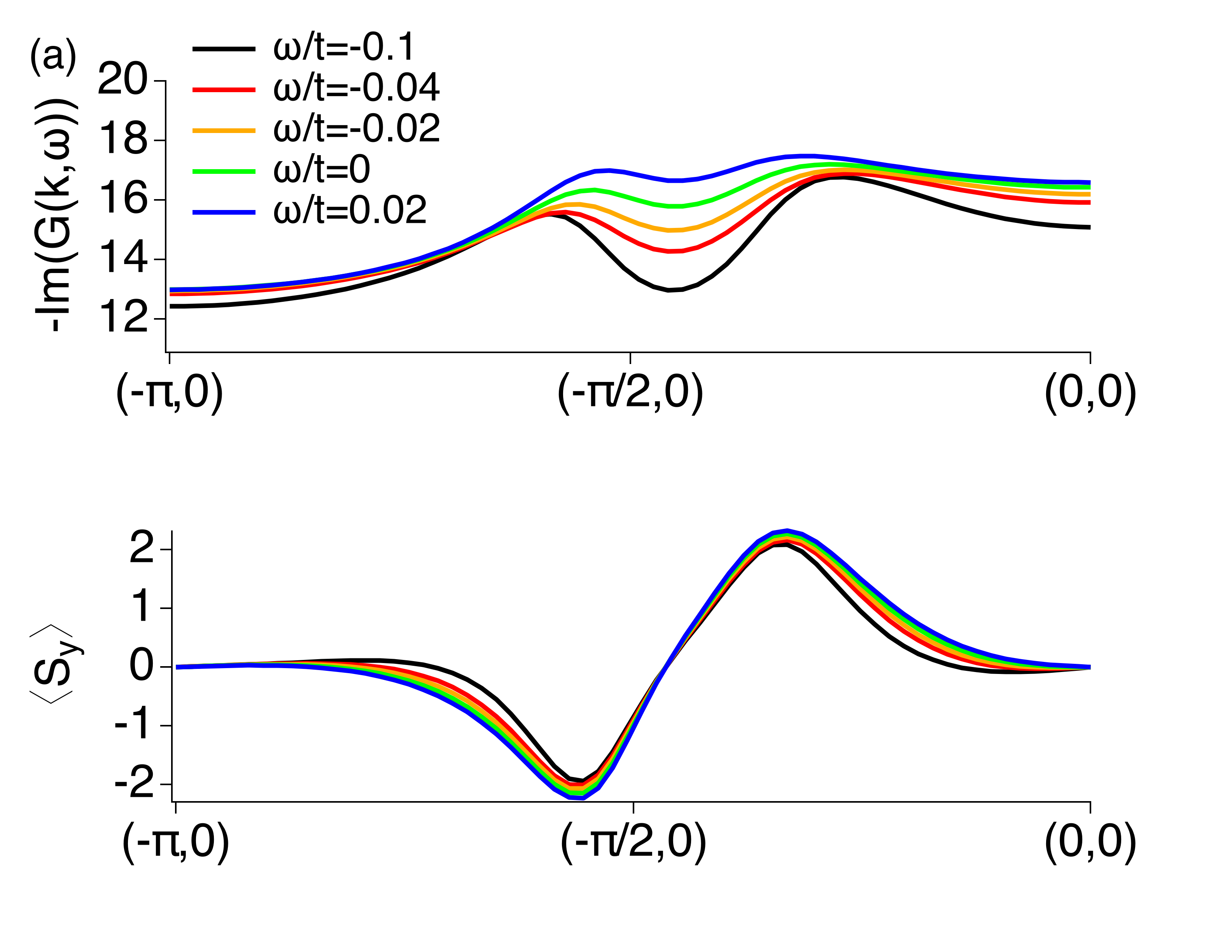}
\includegraphics[width=0.32\linewidth]{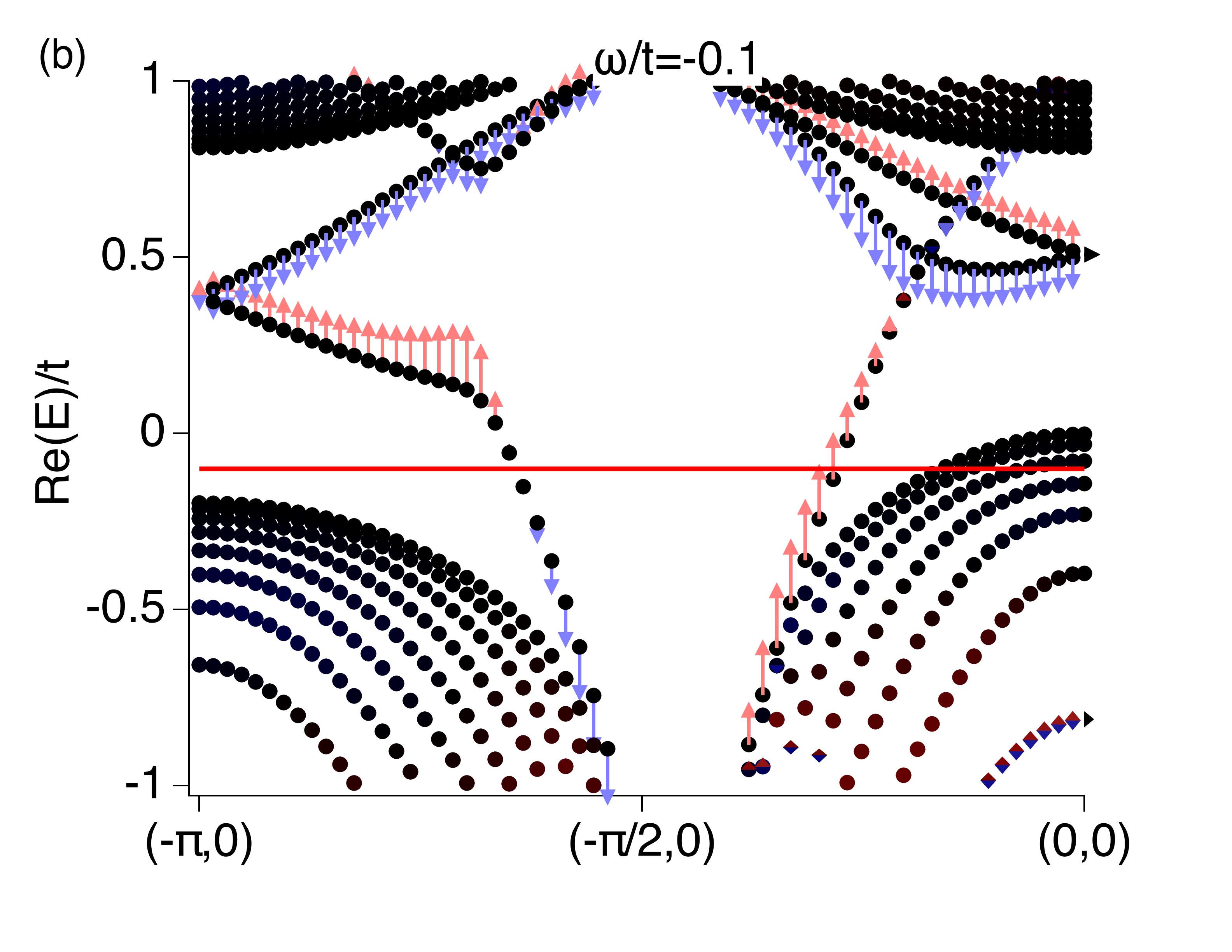}
\includegraphics[width=0.32\linewidth]{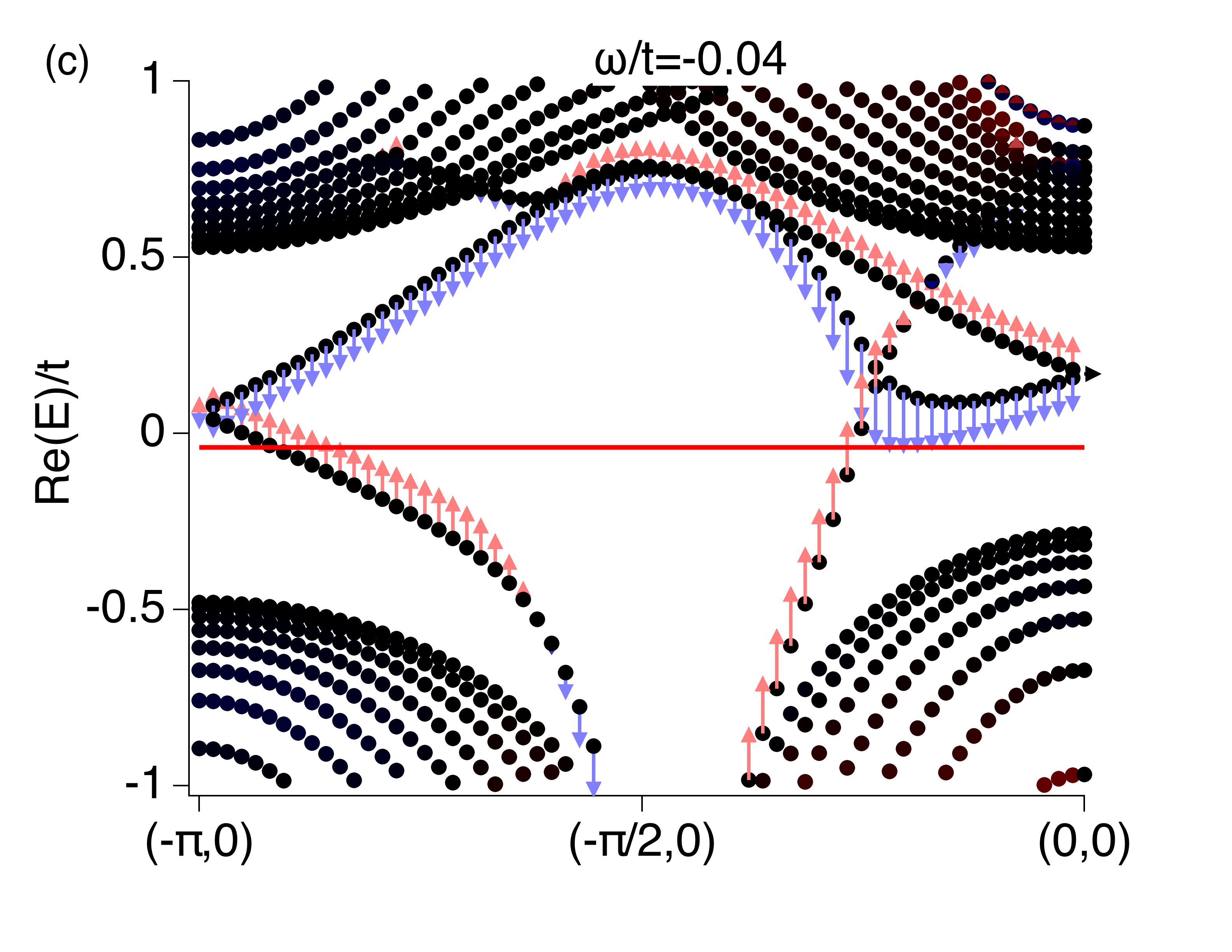}
\includegraphics[width=0.32\linewidth]{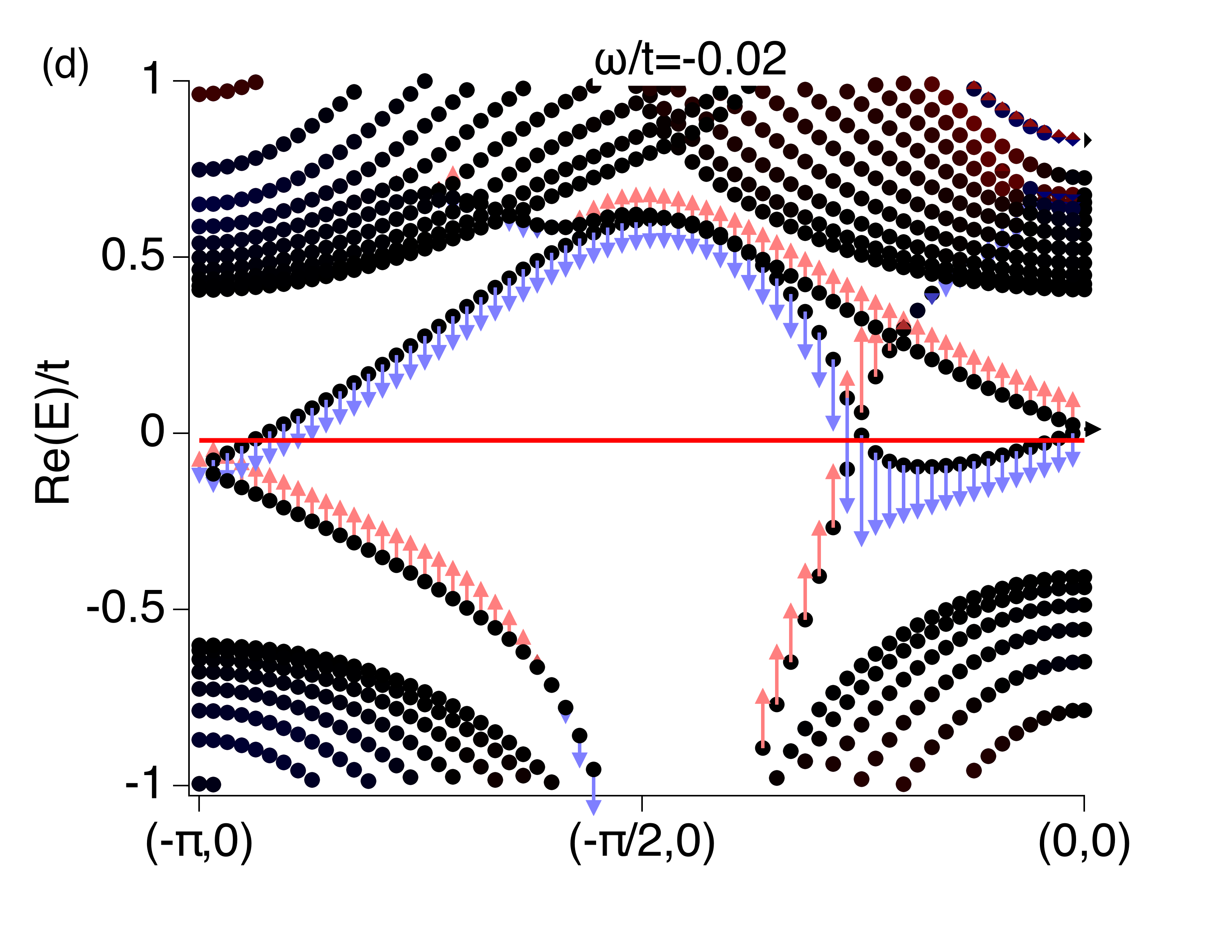}
\includegraphics[width=0.32\linewidth]{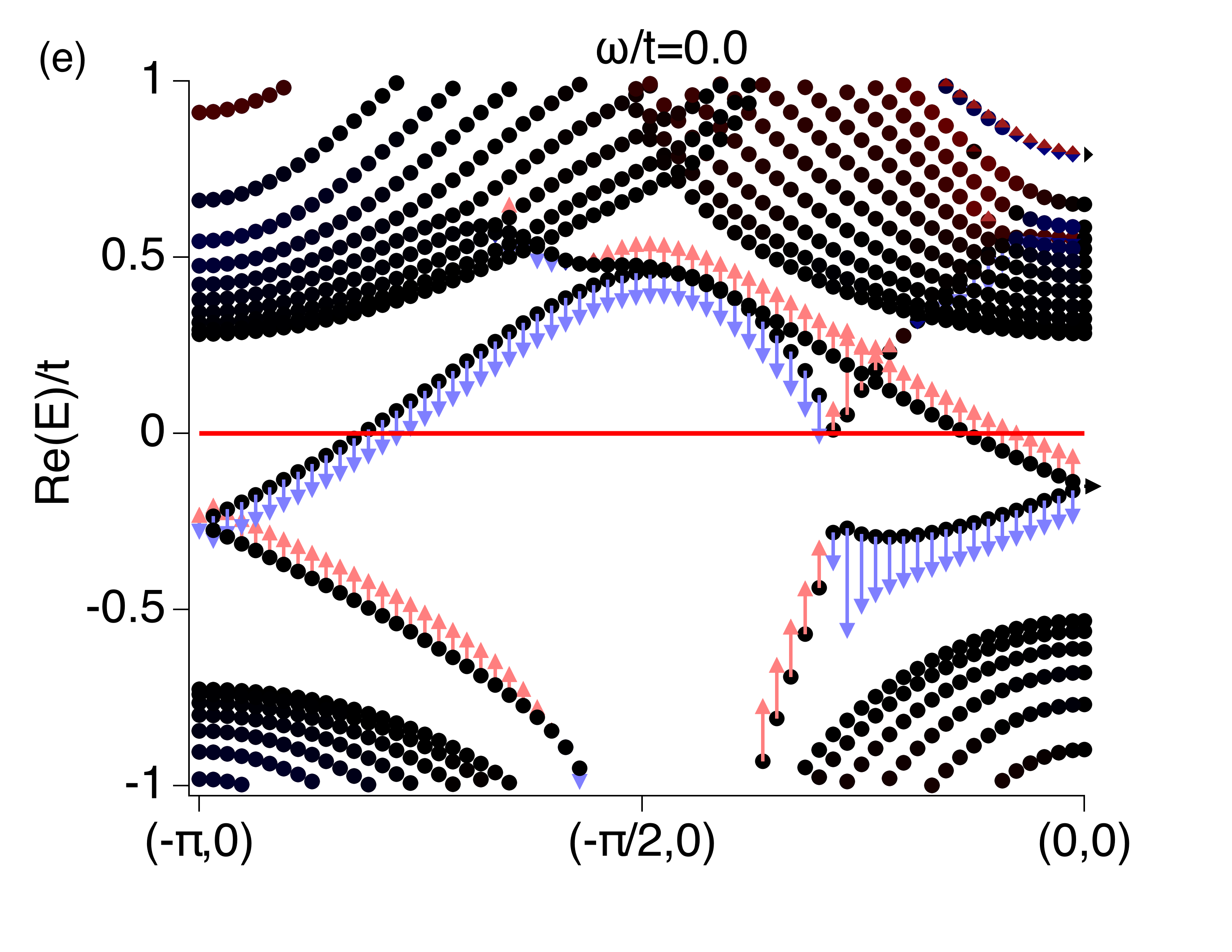}
\includegraphics[width=0.32\linewidth]{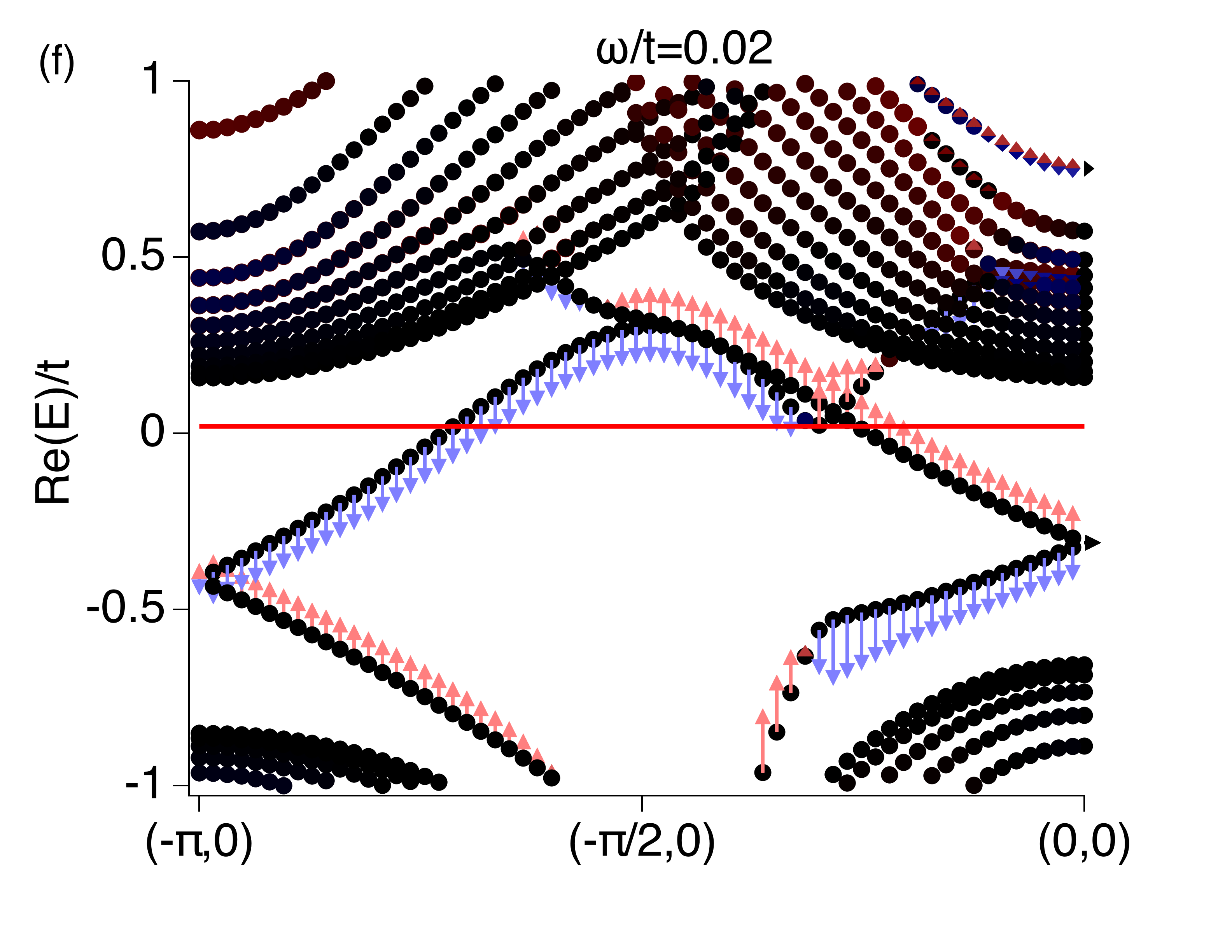}
\caption{Momentum-resolved spectral function and energy spectrum for $T/t=0.106$. Panel (a) shows the spectral function and the spin expectation in the $y$-direction on the bottom surface, $\langle S_y\rangle$, as calculated by the Green's function for different $\omega$ from $(-\pi,0)$ to $(0,0)$ in the BZ. Panels (b)-(f) show the real part of the energies of the effective Hamiltonian for different frequencies along the same path in the BZ. Furthermore, we show the contribution of each state to the spin expectation value,  $\langle S_y\rangle$, of the bottom layer as an arrow attached to each state. The sum of the contributions of all states corresponds to the expectation value shown in panel (a). The frequencies are $\omega/t=-0.1$ (b), $\omega/t=-0.04$ (c),$\omega/t=-0.02$ (d), $\omega/t=0$ (e), and $\omega/t=0.02$ (f).The red lines denote the frequency of the effective Hamiltonian.   \label{Fig12}}
\end{figure*}
Because the self-energy is frequency-dependent, the effective Hamiltonian also possesses a frequency dependence.
We show in Fig.~\ref{Fig12} how the Green's function, the spin expectation value, and the energy spectrum of the effective Hamiltonian change depending on the frequency. The temperature in this figure is fixed to $T/t=0.106$, for which exceptional points at the Fermi energy are absent, as shown in Fig.~\ref{Fig7}. Because the magnitude of the imaginary part of the self-energy increases when changing the frequency away from the Fermi energy (see Fig.~\ref{Fig1}),  exceptional points may emerge in the spectrum away from $\omega=0$. Besides this change in the imaginary part, we note that the real part of the self-energy also varies with frequency. The real part of the self-energy results in a shift of the eigenvalues of the effective Hamiltonian.
 Figure~\ref{Fig12}(a) shows the spectral function and the spin expectation value in the $y$-direction of the bottom layer along a path from $(k_x,k_y)=(-\pi,0)$ to $(0,0)$ in the BZ. In both functions, we see two peaks whose momenta are nearly frequency independent. The spectral function shows only a slight frequency dependence between the maxima. Figures~\ref{Fig12}(b)-(f) show the real part of the eigenvalues of the effective Hamiltonians. We see that these eigenvalues are strongly frequency-dependent and shift when we change the frequency. Although the intersection between the surface states and the red line (frequency of the Hamiltonian) occurs at different momenta, this intersection does not lead to a peak in the spectral function, as shown in Fig.~\ref{Fig12}(a). The reason for this behavior is the imaginary part of the eigenstates present in all surface states at this temperature. Reducing the temperature reduces the imaginary part and thus will lead to visible shifts in the spectral function when we change the frequency. However, in that case, exceptional points do not occur because the imaginary part in the self-energy is necessary for their appearance. The same holds for the spin-expectation value. Because of the imaginary part, all surface states contribute to the spin expectation value.

Finally, let us analyze the emergence of exceptional points in the frequency-dependent effective Hamiltonian. In Fig.~\ref{Fig12}(b), at $\omega/t=-0.1$, we see that two levels cross at $(k_x,k_y)=(-1,0)$ at $\text{Re}(E)/t\approx 0.5$. This level crossing is due to the imaginary part of the self-energy, as has been analyzed in detail in the main text. We see how these two bands separate when increasing the frequency of the effective Hamiltonian from $\omega/t=-0.02$ to $\omega/t=0.0$. Thus, there must be an exceptional point. Again, we note that the changes in the spectral function and the spin expectation value between these two frequencies are small, although an exceptional point appears. Thus, changing the frequency does not provide additional clues to the existence of exceptional points. As described in the main text, the momentum dependence of the spectral function and the spin expectation value at the Fermi energy displaying typical Fermi arc behavior provide better information.

\bibliography{Ref.bib}

\end{document}